\documentclass[print]{aa}

\usepackage{ifpdf}
\ifpdf
\usepackage{graphicx}
\usepackage{hyperref}
\else
\usepackage{graphicx}
\fi

\pdfoutput 0
\usepackage{txfonts}

\title{Direct measurements of the magnification produced by galaxy clusters as gravitational lenses}
\author{Alessandro Sonnenfeld}

\begin{document}
   \title{Direct measurement of the magnification produced by galaxy clusters as gravitational lenses}

   \author{A. Sonnenfeld\inst{1}, G. Bertin\inst{2} \and M. Lombardi\inst{2,3}}

   \institute{Department of Physics, University of California, Santa Barbara, CA 
93106, USA\\
              \email{sonnen@physics.ucsb.edu}
         \and
             Dipartimento di Fisica, Universit\`{a} degli Studi di Milano, via Celoria 16, I-20133 Milan, Italy\\
             \email{giuseppe.bertin@unimi.it}
         \and
         	    European Southern Observatory, Karl-Schwarzschild
Strasse 2, D-85748 Garching, Germany             
             }

   \date{Received Dec 16, 2010; accepted May 29, 2011}

 \abstract
   {Weak lensing is one of the best available diagnostic tools to measure the total density profiles of distant clusters of galaxies. Unfortunately, it suffers from the well-known mass-sheet degeneracy, so that weak lensing analyses cannot lead to fully reliable determinations of the total mass of the clusters. One possible way to set the relevant scale of the density profile would be to make a direct measurement of the magnification produced by the clusters as gravitational lenses; in the past this objective has been addressed in a number of ways, but with no significant success.}
   {In this paper we revisit a suggestion made a few years ago for this general purpose, based on the use of the Fundamental Plane as a standard rod for early-type galaxies. Here we move one step further, beyond the simple outline of the idea given earlier, and quantify some statistical properties of this innovative diagnostic tool, with the final goal of identifying clear guidelines for a future observational test on concrete cases, which turns out to be well within the current instrument capabilities.}
   {The study is carried out by discussing the statistical properties of Fundamental Plane measurements for a sample of early-type source galaxies behind a massive cluster, for which a weak lensing analysis is assumed to be available. Some general results are first obtained analytically and then tested and extended by means of dedicated simulations. }
   {We proceed to study the best strategy to use Fundamental Plane measurements to determine the mass scale of a given cluster and find that the optimal choice is that of a sample of early-type galaxies behind the cluster distributed approximately uniformly in the sky. We discuss the role of the redshift distribution of the source galaxies, in relation to the redshift of the lensing cluster and to the limitations of Fundamental Plane measurements. Simple simulations are carried out for clusters with intrinsic properties similar to those of the Coma cluster. 
We also show that, within a realistic cosmological scenario, substructures do not contribute much to the magnification signal that we are looking for, but only add a modest amount of scatter.}
   {We find that for a massive cluster  ($M_{200} > 10^{15} ~M_{\odot}$) located at redshift $0.3 \pm 0.1$, a set of about 20 Fundamental Plane measurements, combined with a good weak lensing analysis, should be able to lead to a mass determination with a precision of 20 \% or better.}


\titlerunning{Direct measurements of the lensing magnification by galaxy clusters}
\authorrunning{A. Sonnenfeld et al.}
   \maketitle

\section{Introduction}

   Weak lensing is a powerful tool to probe the mass distribution of massive clusters of galaxies.
   Based on the study of the distortion induced by the lens on images of extended background sources, 
   weak lensing techniques have been often used to measure masses of clusters of galaxies (see e.g. Lombardi et al. \cite{Lombardi2000}; Clowe \& Schneider \cite{CloweSchneider}; Broadhurst et al. \cite{Broad05}; Clowe et al. \cite{Bullet}; Gavazzi et al. \cite{Gavazzi09}).
   However, weak lensing suffers from a fundamental limitation set by the so-called mass--sheet degeneracy: the projected surface mass density map $\kappa(\vec{\theta})$ can be determined only up to transformations of the form
   \begin{equation}\label{masssheet}
   \kappa(\vec{\theta}) \mapsto \kappa'(\vec{\theta}) = \lambda\kappa(\vec{\theta}) + 1 - \lambda.
   \end{equation}
In other words, with weak lensing measurements alone it is not possible to constrain the total mass of a lens, unless further assumptions are made.

Although many strategies have been proposed to break this degeneracy, no definitive solution has been found so far.
In principle, the mass-sheet degeneracy can be removed with the determination of the absolute value of $\kappa$ at a single point in the lens plane where $\kappa($\vec{\theta}$) \neq 1$. One can assume that the surface mass density vanishes at the boundaries of the image, far from the lens, and impose that the average value of $\kappa$ along the sides of the field of observation is zero.
However, this assumption requires the field of view to be sufficiently large, which is not always possible. Moreover, current structure formation models predict that many clusters of galaxies have nonvanishing surface mass densities far from the lens center, so that such an assumption is bound to lead to total mass estimates significantly underestimated. Another possibility is to set $\kappa = 0$ at its minimum so that the mass density is everywhere positive, which might seem a plausible assumption. The problem with this approach is that noise can produce negative values of $\kappa$, and adjusting the overall density profile on the basis of a noise feature may not be wise.
A popular solution is to fit weak lensing measurements to parametric model mass distributions, usually NFW (Navarro et al. \cite{NFW}) profiles. This method has the disadvantage that it relies on an assumed mass profile for the lens, which might differ from that of the actual cluster, thus reducing the power of weak lensing as a mass measurement technique.

If we wish to find an assumption-free method to break the mass-sheet degeneracy, additional information must be added to the weak lensing measurements.
Several attempts have been made in this direction.
One possibility is to include magnification information in the data set.
In fact, measurements of the magnification and of the shear field can lead to a direct determination of the surface mass density $\kappa$.
Broadhurst et al. (\cite{Broad95}) proposed a method based on the study of the number counts of faint background galaxies for the determination of the magnification. Their technique was successfully used in a few cases of particularly massive clusters (Fort et al. \cite{Fort}, Taylor et al. \cite{Taylor}, Broadhurst et al. \cite{Broad05}, Umetsu et al. \cite{Umetsu}). However, for this method a detailed calibration of nontrivial model quantities, such as the number counts of unlensed sources, is essential. Moreover, the count process in the central regions of rich clusters is made difficult by bright cluster members that hide the faintest background galaxies.

The form of the invariance transformation (\ref{masssheet}) is referred to a fixed source redshift: the same transformation referred to a source at a different redshift changes through a coefficient that multiplies the term $1 - \lambda$.
In other words, each portion of the redshift space suffers from a different invariance transformation, so that in principle the mass--sheet degeneracy can also be broken by combining lensing information from images of sources at different {\em known} redshifts.
Brada\u{c} et al. (\cite{Bradac04}) investigated this possibility in the context of weak lensing, considering the hypothetical case in which the individual redshifts of the lensed background galaxies are available.
They showed that, under these favourable circumstances, the mass-sheet degeneracy can be broken for critical clusters (i.e. those that can produce multiple images), but still not for subcritical ones.

A significant improvement can come from the addition of information from strongly lensed images (i.e. arcs or multiple images), provided that the redshifts of the strongly lensed sources differ from the mean redshift of the sources used for the weak lensing analysis.
Indeed, several methods based on the inclusion of strong lensing information have been devised and applied to real cases (Brada\u{c} et al. \cite{Bradac05a}, \cite{Bradac05b}, Cacciato et al. \cite{Cacciato}, Diego et al. \cite{Diego}, Merten et al. \cite{Merten}). 
However, strongly lensed images can only be produced by critical clusters.
Thus, for subcritical clusters the mass-sheet degeneracy still remains a fundamental problem in the determination of the total mass.

It is in this context that Bertin \& Lombardi (2006, \cite{BL06} from now on) proposed a new method to measure the lensing magnification induced by a cluster, which can be used to break the mass-sheet degeneracy.
They showed that estimates of the magnification can be obtained by observing background early-type galaxies. Early-type galaxies can be treated as standard rods, in virtue of the empirical law of the Fundamental Plane. From a Fundamental Plane measurement, the intrinsic effective radius of an early-type galaxy can be recovered with a 15\% accuracy (\cite{BL06}).
Then, by measuring the observed (magnified) effective radius, the magnification can be derived.

In this paper we investigate further the possibilities opened by this technique.
In particular, we wish to determine which accuracy on the determination of the total mass of a cluster can be achieved by combining Fundamental Plane measurements with a weak lensing study.
This is done by prescribing a method to break the degeneracy and by studying its properties, both analytically and with the aid of simulations. 
In principle, substructures in the lensing cluster can introduce noise in the magnification signal, so that a given measurement might be used to set interesting constraints on the amount of substructure rather than on the mass of the cluster.
This issue is also studied in this paper, with the use of numerical simulations.
Then we address the problem of identifying the optimal conditions on lens and source-galaxies in order to get a significant measurement in concrete cases.

The structure of the paper is the following. In Sect.~\ref{section2} we give the basic lensing equations and present the problem of the mass-sheet degeneracy. In Sect.~\ref{method} we show how Fundamental Plane measurements can be used to infer the magnification and present a simple method to use such measurements to break the mass--sheet degeneracy. In Sect.~\ref{statistics} the statistical properties of this method are studied. In Sect.~\ref{substructures} we address the issue of how substructures can influence the magnification signal we seek to measure. In Sect.~\ref{Simulations} we describe the simulations set up to test the method and show the results. Conclusions are drawn in Sect.~\ref{Conclusions}.

\section{Weak lensing preliminaries}\label{section2}

\subsection{Basic notation and equations}

We start by introducing the projected surface mass density of a given lens, $\Sigma(\vec{\theta})$.
The nature of the lensing equations make it convenient to introduce the {\em dimensionless surface mass density} $\kappa(\vec{\theta},z)$ for a source at redshift $z$, defined as
\begin{equation}
\kappa(\vec{\theta},z) \equiv \frac{\Sigma(\vec{\theta})}{\Sigma_{cr}(z)}\quad{\rm where}\quad\Sigma_{cr}\equiv \frac{c^2}{4\pi G}\frac{D_s}{D_dD_{ds}}.
\end{equation}
Here $D_d$, $D_s$, $D_{ds}$ are angular diameter distances of the lens and source with respect to the observer, and of the source with respect to the lens, respectively.
The surface mass density is referred to a fiducial source at infinite redshift,
\begin{equation}
\kappa(\vec{\theta},z) \equiv Z(z)\kappa(\vec{\theta}),
\end{equation}
through the cosmological weight function
\begin{equation}
Z(z) \equiv \frac{\lim_{z' \rightarrow \infty} \Sigma_{cr}(z')}{\Sigma_{cr}(z)}H(z-z_d).
\end{equation}
Here $H(z-z_s)$ is the Heaviside step function, to take into account that images of sources at redshift lower than that of the lens are not lensed.

An important quantity that enters the weak lensing problem is the redshift--dependent reduced shear $g(\vec{\theta},z)$:
\begin{equation}
g(\vec{\theta},z) = \frac{Z(z)\gamma(\vec{\theta})}{1 - Z(z)\kappa(\vec{\theta})},
\end{equation}
where  the {\em shear} $\gamma(\vec{\theta})$ can be expressed as a nonlocal function of $\kappa(\vec{\theta})$:
\begin{equation}\label{gammakappa}
\gamma(\vec{\theta}) = \frac{1}{\pi}\int_{\mathbb{R}^2}\mathcal{D}(\vec{\theta} -\vec{\theta}')\kappa(\vec{\theta}')d^2\theta',
\end{equation}

\begin{equation}
\mathrm{with}\quad \mathcal{D}(\vec{\theta}) \equiv -\frac{1}{(\theta_1 - i\theta_2)^2}.
\end{equation}
Weak lensing consists in the study of the distortion induced by the lens on the images of background sources. 
This can be done in terms of a {\em complex ellipticity} $\epsilon$, defined from the quadrupole moments $Q_{ij}$ of the surface brightness as
\begin{equation}
\epsilon \equiv \frac{Q_{11} - Q_{22} + 2iQ_{12}}{Q_{11}+Q_{22} + 2\sqrt{Q_{11}Q_{22} - Q_{12}^2}}.
\end{equation}
Seitz \& Schneider (\cite{SS97}) showed that the image ellipticity is related to the intrinsic (unlensed) source ellipticity $\epsilon^s$ in the following way:
\begin{equation}\label{ellipticity}
\epsilon = \left\{\begin{tabular}{ll} $\displaystyle\frac{\epsilon^s + g(\vec{\theta},z)}{1 + g(\vec{\theta},z)^*\epsilon^s}$ & if $|g(\vec{\theta},z)| < 1$ \\
 & \\
$\displaystyle\frac{1 + g(\vec{\theta},z)\epsilon^{s*}}{\epsilon^{s*} + g(\vec{\theta},z)^*}$ & otherwise \end{tabular}\right. .
\end{equation}
Therefore, under the assumption that the intrinsic ellipticity distribution of background sources is isotropic, the expectation value of the observed ellipticity for sources at redshift $z$ is:
\begin{equation}
\mathrm{E}[\epsilon(z)] = \left\{\begin{tabular}{ll} $g(\vec{\theta},z)$ & if $|g(\vec{\theta},z)| < 1$ \\
&\\
$\displaystyle \frac{1}{g^*(\vec{\theta},z)}$ & otherwise \end{tabular}\right.
\end{equation}
In the general case of sources distributed in redshift, the following approximation holds for $\kappa \lesssim 0.6$:
\begin{equation}\label{expect}
\mathrm{E}[\epsilon] \simeq \frac{\left<Z\right>\gamma(\vec{\theta})}{1-\frac{\left<Z^2\right>}{\left<Z\right>}\kappa(\vec{\theta})},
\end{equation}
as shown by Seitz \& Schneider (\cite{SS97}). $\left<Z^n\right>$ are the moments of the redshift probability distribution of the background galaxies.

\subsection{The mass--sheet degeneracy}

By averaging over image ellipticities of background galaxies and identifying $\left<\epsilon\right>$ with $\mathrm{E}[\epsilon]$, we can estimate the quantity (\ref{expect}) in the field of observation.
Then, since $\gamma$ depends on $\kappa$ through Eq. (\ref{gammakappa}), relation (\ref{expect}) can be inverted and the surface mass density $\kappa$ can be recovered from the observed average ellipticity.
Practical realizations of this picture were provided by Kaiser (\cite{Kaiser95}), Seitz \& Schneider (\cite{SS97}), Lombardi \& Bertin (\cite{Lombardi1999}).

It can be shown that the quantity $\mathrm{E}[\epsilon]$ in Eq. (\ref{expect}), which is the observable quantity, is invariant under transformations of the form
\begin{equation}\label{invariance}
\kappa(\vec{\theta}) \rightarrow \kappa'(\vec{\theta}) = \lambda\kappa(\vec{\theta}) + w(1 -\lambda),
\end{equation}
where we introduced
\begin{equation}
w \equiv \frac{\left<Z\right>}{\left<Z^2\right>}.
\end{equation}
This is the mass--sheet degeneracy for the general case of sources distributed in redshift: with ellipticity measurements it is only possible to recover the surface mass density up to the above transformation.
In the simplified case of sources all at the same redshift $z$, the invariance transformation reduces to (\ref{masssheet}), with $\kappa(\theta,z) = \kappa(\vec{\theta})$.

In principle, the mass--sheet degeneracy can be broken with a local measurement of the magnification. In fact, the magnification $\mu(\vec{\theta},z)$ is related to $\kappa(\vec{\theta})$ and $\gamma(\vec{\theta})$ through the following relation:
\begin{equation}\label{magnification}
\mu(\vec{\theta},z) = \left|[1 - Z(z)\kappa(\vec{\theta})]^2 - Z^2(z)|\gamma(\vec{\theta})|^2\right|^{-1}.
\end{equation}
By measuring $\left<\epsilon\right>$ and $\mu(z)$ and by combining (\ref{expect}) with (\ref{magnification}) we obtain a two--equation system in terms of $\kappa$ and $\gamma$, which can be solved for $\kappa$, thus breaking the mass--sheet degeneracy. This requires that we know which redshift $z$ the magnification measurement is referred to, in order to calculate the cosmological weight $Z(z)$ that enters Eq. (\ref{magnification}).

\section{Breaking the mass--sheet degeneracy}\label{method}

In this section we will introduce a mass measurement method based on the combined use of weak lensing and magnification measurements.

\subsection{The Fundamental Plane}

The Fundamental Plane (FP from now on; Dressler et al. \cite{Dressler}, Djorgovski \& Davis \cite{Djorgovski}) is an empirical scaling law that applies to early-type galaxies (E/S0). It relates three well-defined observable quantities for these objects: the {\em effective radius}, $R_e$, the {\em effective surface brightness}, $\left<\mathrm{SB}\right>_e$, and the {\em central velocity dispersion} of the stellar component, $\sigma_0$.
The three quantities are related in the following way:
\begin{equation}\label{FundamentalPlane}
\mathrm{Log}\, R_e = \mathrm{Log}\,r_e + \mathrm{Log}\,D_A(z) = \alpha \mathrm{Log}\,\sigma_0 + \beta \left<\mathrm{SB}\right>_e + \gamma,
\end{equation}
where $\alpha$, $\beta$, and $\gamma$ are empirically determined coefficients that depend on the waveband of observation, $r_e$ is the effective radius in angular units and $D_A(z)$ is the angular diameter distance of the galaxy at redshift $z$.
The existence of such a relation has been extensively confirmed by a number of studies on both field and cluster galaxies out to cosmological distances (e.g. J\o rgensen et al. \cite{Jorgensen93}, Bender et al. \cite{Bender98}, Treu et al. \cite{Treu99}). 
The measurement of the Fundamental Plane parameters for the most distant sample of objects has been carried out by van der Wel et al. (\cite{vanderWel}, vdW05 from now on), who examined early-type galaxies out to $z \approx 1.1$.
This relation is observed to hold within a $0.07$ scatter on ${\rm Log}\, r_e$, or $15 \%$ on $r_e$, rather independently of the position on the FP plane (J\o rgensen et al. \cite{Jorgensen96}) and increasing with increasing redshift (Treu et al. \cite{Treu05}). 
Treu et al. (\cite{Treu05}) have quantified the increase in the scatter in ${\rm Log}\,r_e$ of the FP relation as $d\sigma_\gamma/dz = 0.032 \pm 0.012$, which translates into a scatter of $23\%$ in $r_e$ at $z = 1$.
It is still not clear if the source of this scatter is totally intrinsic or if it can be reduced by improving the observational precision.
Auger et al. (\cite{Auger}) estimated the intrinsic scatter of the FP to be as low as 11\%.

Observations have pointed out a variation of the coefficient $\gamma$ with redshift, quantified as $d\gamma/dz = 0.58_{-0.06}^{+0.04}$ by Treu et al. (\cite{Treu05}). For cluster galaxies there is evidence for a slower evolution (see e.g. Wuyts et al. \cite{Wuyts}, vdW05), so that the Fundamental Plane for cluster galaxies appears to differ from that of field galaxies. In vdW05 it was shown that this difference is not significant for massive ($M > 2\times 10^{11} M_\odot$) galaxies, and a similar result was found by van Dokkum \& van der Marel (\cite{vv07}). According to vdW05, the scatter of the FP is also smaller for the more massive objects. Evidence for a variation of the coefficients $\alpha$ and $\beta$ with redshift has also been reported (Treu et al. \cite{Treu05}), but this is generally taken to be less significant.

\subsection{The FP seen through a lens}

As shown in \cite{BL06}, the Fundamental Plane changes in a well defined way when viewed through a gravitational lens.
Both the surface brightness $\left<\mathrm{SB}\right>_e$ and the central velocity dispersion $\sigma_0$ are lens-invariant.
Therefore, by measuring these two quantities for early-type galaxies and by making use of the Fundamental Plane relation (\ref{FundamentalPlane}) it is possible to obtain an estimate of the effective radius of the observed galaxy, $R_e^{\mathrm{(FP)}}$.
By measuring the redshift of the galaxy it is then possible to convert this measurement of the effective radius to angular units, $r_e^{\mathrm{(FP)}}$.
By observing the effective radius $r_e^{\mathrm{(obs)}}$, magnified by the lens effect, the lens magnification will then be given by the square of the ratio of the intrinsic size to the observed image size, as
\begin{equation}
\mu = \left(\frac{r_e^{\mathrm{(obs)}}}{r_e^{\mathrm{(FP)}}}\right)^2.
\end{equation}
With a 15\% scatter of the Fundamental Plane relation, the same error will affect our estimate of the intrinsic effective radius, $r_e^{\mathrm{(FP)}}$, which translates into a $\sim 30\%$ error on the magnification.

Actually, the quantity that is lens invariant is the intrinsic surface brightness, which in general differs from the observed surface brightness.
Nevertheless, van der Wel et al. (\cite{vanderWel}) showed that the intrinsic surface brightness of galaxies at high redshift can be effectively measured by fitting Sersic models convolved with the PSF. We expect this task to be made easier by the magnifying effect of lensing.

\subsection{A minimum--$\chi^2$ approach}

The method presented here has been developed and tested for noncritical lenses only, although it can be generalized to the critical case. 
Therefore, from now on we will assume that the lens does not have critical curves, unless stated differently. This is also the most interesting case, because it is for subcritical lenses that the problem of the mass-sheet degeneracy is harder to overcome.

Suppose we performed a weak lensing analysis of a cluster, which led to the determination of the average distortion $\left<\epsilon\right>(\vec{\theta})$ within the field of view. By inverting the distortion map (for instance with the nonparametric method of Lombardi \& Bertin \cite{Lombardi1999}) it is possible to recover the surface mass density of the lens $\kappa(\vec{\theta})$ up to the invariance transformation (\ref{invariance}).
Then, suppose that we performed a set of $N_{\mathrm{FP}}$ Fundamental Plane measurements, through which we estimate the magnification in a corresponding number of positions $\{\vec{\theta}_i\}$ on the image plane. 

Among the infinite possible mass density maps compatible with the distortion measurements, spanned by (\ref{invariance}), we select the one for which the accordance with the observations of magniÞed early-type galaxies is best.
To do so, we first transform the estimates of the magnification into estimates of the surface mass density, $\kappa^{\mathrm{(FP)}}(\vec{\theta}_i)$, making use of Eqs. (\ref{expect}) and (\ref{magnification}).
This requires that we know the average distortion in the image position, $\left<\epsilon\right>(\vec{\theta}_i)$, determined from the weak lensing study, and the redshift $z_i$ of each galaxy, which can be measured with the same spectroscopic observation necessary to measure $\sigma_0$.
The expression that gives $\kappa^{\mathrm{(FP)}}(\vec{\theta}_i)$ in terms of $\mu(\vec{\theta}_i)$, $\left<\epsilon\right>(\vec{\theta}_i)$ and $z_i$ is the following:
\begin{equation}\label{kappaFP}
\kappa^{\mathrm{(FP)}} = \frac{-b-\sqrt{b^2 - a(c - 1/\mu)}}{a},
\end{equation}
where
\begin{equation}\label{abkappa}
a \equiv Z_i^2\left(1 - \frac{1}{w^2\left<Z\right>^2}|\left<\epsilon\right>|^2\right),\quad b\equiv Z_i\left(\frac{Z_i}{w\left<Z\right>^2}|\left<\epsilon\right>|^2 - 1\right),
\end{equation}
\begin{equation}\label{ckappa}
c \equiv 1 - \frac{Z_i^2}{\left<Z\right>^2}|\left<\epsilon\right>|^2 \qquad \mathrm{and}\quad Z_i \equiv Z(z_i).
\end{equation}
The minus sign for the square root in (\ref{kappaFP}) comes from the fact that the lens was assumed to be everywhere subcritical.

The fit is performed by minimizing the following penalty function:
\begin{equation}\label{chisq}
\chi^2 = \sum_{i=1}^{N_{\mathrm{FP}}}\frac{1}{\sigma_i^2}\left|\lambda\kappa_0(\vec{\theta}_i) + w(1 - \lambda) - \kappa_i^{\mathrm{(FP)}}\right|^2,
\end{equation}
where $\kappa_0(\vec{\theta})$ is the surface mass density distribution inferred from the weak lensing reconstruction, and $\sigma_i^2$ are suitably chosen weights.
The minimum--$\chi^2$ condition can be found by imposing $\partial \chi^2/\partial\lambda = 0$, which leads to the determination of the estimator $\hat{\lambda}$ for the parameter $\lambda$:
\begin{equation}\label{estimator}
\hat{\lambda} = \frac{\displaystyle \sum_{i=1}^{N_{\mathrm{FP}}}\frac{1}{\sigma_i^2}[\kappa_i^{\mathrm{(FP)}} - w][\kappa_0(\vec{\theta}_i)-w]}{\displaystyle \sum_{i=1}^{N_{\mathrm{FP}}}\frac{1}{\sigma_i^2}[\kappa_0(\vec{\theta}_i) - w]^2}.
\end{equation}
The choice of the weights $\sigma_i^2$ that enter this $\chi^2$ function requires particular care.
Generally, in minimum-$\chi^2$ approaches the values of $\sigma_i$ are taken to be proportional to the measurement errors of the quantity over which the fit is performed: in the present case, the surface mass density $\kappa^{\mathrm{(FP)}}$.
By assuming that the errors come only from the Fundamental Plane measurements (i.e. under the assumption of perfect weak lensing measurements), error propagation on $\kappa^{\mathrm{(FP)}}$ gives
\begin{equation}\label{errork}
\Delta\kappa_{\mathrm{FP}} = \sigma_{\mathrm{FP}}(z) \frac{\displaystyle [1 - Z(z)\kappa]^2 - \frac{Z(z)^2}{\left<Z\right>^2}\left(1 - \frac{\kappa}{w}\right)^2|\left<\varepsilon\right>|^2(\vec{\theta})}{\displaystyle Z(z)[1 - Z(z)\kappa] - \frac{Z(z)^2}{w\left<Z\right>^2}\left(1 - \frac{\kappa}{w}\right)^2|\left<\varepsilon\right>|^2(\vec{\theta})},
\end{equation}
where $\sigma_{\mathrm{FP}}(z)$ is the scatter in $r_e$ of the Fundamental Plane.
Therefore, $\Delta\kappa_{\mathrm{FP}}$ depends on the value of $\kappa$, which is the quantity we are trying to determine with the fit. This complicates the definition of the weights.
One would be tempted to define $\sigma_i = \Delta\kappa(\kappa^{\mathrm{(FP)}}(\vec{\theta}_i))$, but by doing so we would introduce a bias in the estimate of $\lambda$: 
this is because a higher weight would be given to measurements in which fluctuations give higher values of the magnification, relative to cases in which $\mu$ is underestimated. Then the fit would yield preferentially higher mass estimates.

\begin{figure}
\setlength{\unitlength}{1cm} 
\begin{picture}(6,5)
\thicklines
\put(4,4.5){\fbox{Weak lensing reconstruction}}
\put(6,4){\vector(0,-1){1}}
\put(4,2.5){\fbox{$\sigma_i = \Delta\kappa_{\mathrm{FP}}(\kappa^{(n)},|\left<\varepsilon\right>|,z_i)$}}
\put(6,2.2){\vector(0,-1){1.2}}
\put(4.8,3.5){$n=0$}
\put(5.5,0.5){\fbox{$\kappa^{(n+1)}$}}
\put(5.2,0.6){\line(-1,0){2.5}}
\put(2.7,0.6){\line(0,1){2}}
\put(2.7,2.6){\vector(1,0){1}}
\put(0.8,1.6){Iteration}
\put(0.8,1.1){$n = n+1$}
\put(6.5,1.6){Fit}
\end{picture}
\caption{The fitting process.}\label{schemino}
\end{figure}
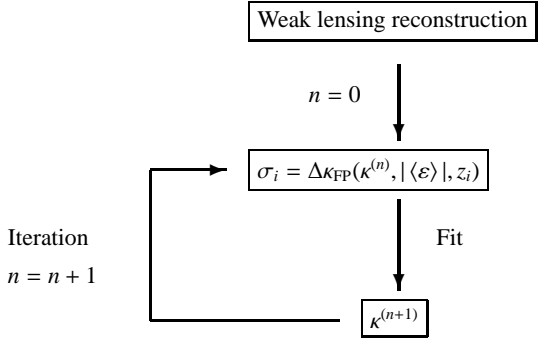

The solution we propose is an iteration procedure, seeded by the definition of the weights based on the model surface mass density $\kappa_0$ obtained from weak lensing: in (\ref{errork}) we identify $\kappa_0$ with $\kappa$ and define
\begin{equation}
\sigma_i \equiv \Delta\kappa_{\mathrm{FP}}(\kappa_0(\vec{\theta}_i)).
\end{equation}
We can then minimize the $\chi^2$ function and obtain a new estimate of the density map. If $\lambda_{(0)}$ is the value of the parameter $\lambda$ obtained from the fit with (\ref{estimator}), then this new model for the surface mass density will be given by
\begin{equation}
\kappa_1 = \lambda_{(0)}\kappa_0 + w(1-\lambda_{(0)}).
\end{equation}
At this point we can update the weights to this new model by redefining $\sigma_i = \Delta\kappa_{\mathrm{FP}}[\kappa_1(\vec{\theta}_i)]$ and iterate the procedure.
Fig. \ref{schemino} outlines the basic steps of the method.
The method converges quickly and the final solution is found to be invariant under transformations of the form (\ref{invariance}) of the input surface mass density map $\kappa_0$.

Eq. (\ref{chisq}) is only one possible definition of a $\chi^2$ function for our problem. We could have defined a $\chi^2$ in terms of other quantities, such as the magnification $\mu$ or its inverse $|\det A|$. The advantage of our choice lies in the fact that the minimum $\chi^2$ condition $\partial\chi^2/\partial\lambda = 0$ is a linear function of $\lambda$. This makes it relatively easy to study the statistical properties of the method, which will be the subject of Sect.~\ref{statistics}.

\section{Statistical properties of the method}\label{statistics}

To assess the reliability of the method introduced to break the mass-sheet degeneracy, it is necessary to study in detail the statistical properties of the results obtained.
In particular, we wish to clarify if the estimate of the total mass resulting from this procedure is biased, and to determine the expected uncertainty.
%
%
%
%
%
%
%
%

In order to do this, we need to discuss the conditions of observation first.
Obviously, it would be preferable to work with a large number of measurements, but a practical limit is imposed by the telescope time required to measure velocity dispersions.
We believe that a reasonable number of Fundamental Plane measurements that can be effectively carried out in an observational campaign is $N_{\mbox{FP}} \approx 20$. 
Another important issue is the possibility of Þnding a sufficient number of sources behind the observed gravitational lens.
\cite{BL06} estimated the number density of practically observable sources to be $\sim 2\,\mathrm{arcmin}^{-2}$. This estimate was based on a study of magnitude--number counts of early--type galaxies (Glazebrook et al. \cite{Glazebrook}).
This sets a limit on the field of view required to have a sufficiently high statistics of Fundamental Plane measurements.
It also tells us that, unless a large field of view is examined, the observer does not have much freedom in selecting objects.
Our statistical study will take into acocunt these important factors.

In this section, the weak lensing study will be assumed to provide a perfect reconstruction of the surface mass density map, except for the mass-sheet degeneracy.
Since the fit method is invariant under tranformations of the form (\ref{invariance}) on the input density map $\kappa_0(\vec{\theta})$, we can also assume without further restrictions that $\kappa_0(\vec{\theta})$ is the exact surface mass density of the lens:
\begin{equation}
\kappa_0(\vec{\theta}) = \kappa_{\mathrm{true}}(\vec{\theta}) \equiv \kappa(\vec{\theta}).
\end{equation}
Based on these assumptions, we will study the statistical properties of the estimator $\hat{\lambda}$ at the first iteration (i.e. $\hat{\lambda}_{(0)}$).
The results obtained will provide information on the error in the determination of the mass density map $\kappa_1(\vec{\theta})$ after the first iteration. Since the method converges quickly, there is little difference between $\kappa_1(\vec{\theta})$ and the final density map, therefore we will consider the difference between $\kappa_1$ and the exact density map $\kappa$ to evaluate errors.

\subsection{Error on $\hat{\lambda}$}

We define the expectation value of the estimator $\hat{\lambda}$ as
\begin{equation}\label{aspettalambda}
{\rm E}\{\hat{\lambda}\} = \sum_{i=1}^{N_{\mathrm{FP}}} {\rm E}\left\{I_i\right\},
\end{equation}
where we introduced
\begin{equation}
I_i\left(\kappa_i^{\mathrm{(FP)}},\{\vec{\theta}_j\},\{z_j\}\right) \equiv \frac{\displaystyle \frac{1}{\sigma_i^2}[\kappa_i^{\mathrm{(FP)}} - w][\kappa(\vec{\theta}_i)-w]}{\displaystyle \sum_{j=1}^{N_{\mathrm{FP}}}\frac{1}{\sigma_j^2}[\kappa(\vec{\theta}_j) - w]^2}.
\end{equation}
%
%
%
%
%
%
%
%
We assume that the number of Fundamental Plane measurements $N_{\mbox{FP}}$ is fixed, and that the image positions are randomly distributed.
%
%
%
%
%
%
%
%
%
%
%
%
%
%
%
%
%
%
We will denote by $P_{\vec{\theta},z}(\vec{\theta},z)$ the probability distribution in redshift and image position of the observed galaxies.

Given these definitions, we write the quantities ${\rm E}\{I_i\}$ that enter (\ref{aspettalambda}) as
%
%
%
%
%
%
%
%
%
\begin{equation}\label{integrali}
{\rm E}\{I_i\} = \prod_{j=1}^{N_{\mathrm{FP}}} \int  P_{\vec{\theta},z}(\vec{\theta}_j,z_j)d\vec{\theta}_jdz_j\int P_\kappa(\kappa_i^{\mathrm{(FP)}})d\kappa_i^{\mathrm{(FP)}} I_i,
\end{equation}
where $P_\kappa$ is the probability distribution for the estimates of $\kappa$ from a Fundamental Plane measurement.
Let us make the assumption that $P_\kappa(\kappa_i^{\mathrm{(FP)}})$ is a Gaussian function with dispersion $\sigma_i$, centered on the exact value of the surface mass density:
\begin{equation}\label{Gaussiandist}
P_\kappa\left(\kappa_i^{\mathrm{(FP)}},\{\kappa,\sigma_i\}\right) = \frac{1}{\sqrt{2\pi\sigma_i^2}}\exp\left\{-\frac{[\kappa_i^{\mathrm{(FP)}} - \kappa(\vec{\theta}_i)]^2}{2\sigma_i^2}\right\}.
\end{equation}
The validity of this assumption will be discussed in Appendix A. We then perform the integration in $d\kappa_i^{\mathrm{(FP)}}$ in (\ref{integrali}), which yields
\begin{equation}
\int d\kappa_i^{\mathrm{(FP)}}P_\kappa(\kappa_i^{\mathrm{(FP)}}) I_i = \frac{\displaystyle \frac{1}{\sigma_i^2}[\kappa(\vec{\theta}_i)-w]^2}{\displaystyle \sum_{j=1}^{N_{\mathrm{FP}}}\frac{1}{\sigma_j^2}[\kappa(\vec{\theta}_j) - w]^2},
\end{equation}
from which, together with (\ref{aspettalambda}), it easily follows that
\begin{equation}\label{nobiassulambda}
{\rm E} \{\hat{\lambda}\} = 1.
\end{equation}
Since the input density map was assumed to be the exact surface mass density of the lens ($\kappa_0(\vec{\theta}) = \kappa(\vec{\theta})$), this result implies that the expectation value for the reconstructed surface mass density distribution is the exact one: {\em the estimator $\hat{\lambda}$ is not biased}. 

As a second step, we study the second moment of the probability distribution for $\hat{\lambda}$, the variance: ${\rm Var}(\hat{\lambda}) = {\rm E}\{\hat{\lambda}^2\} - {\rm E}^2\{\hat{\lambda}\}$. 
A similar calculation leads to 
\begin{equation}\label{varianzalambda}
{\rm Var}(\hat{\lambda}) = {\rm E}\{\hat{\lambda}^2\} - {\rm E}^2\{\hat{\lambda}\} = \left<\frac{1}{\displaystyle \sum_i^{N_{\mathrm{FP}}}\frac{1}{\sigma_i^2}[\kappa(\vec{\theta}_i)-w]^2}\right>,
\end{equation}
where angle brackets indicate that the expression must be averaged over the possible image positions and source redshifts.

We note that in the limit of high $N_{\mathrm{FP}}$ the variance has the typical behavior $\sim 1/{N_{\mathrm{FP}}}$. 
We have exactly ${\rm Var}(\hat{\lambda}) \propto 1/{N_{\mathrm{FP}}}$ in the case of a uniform sheet of constant surface mass density.
%
%
%
%
%
%
%
%
%
%
%
%
%
%
%
%
%
%

\subsection{Error on the total mass}

We have just shown the basic results on the accuracy of the method in determining the parameter ${\lambda}$.
Now we will examine the problem of how an error on $\lambda$ translates into errors on the estimated total mass $M$ of the lens.

The total mass inside the field of view is given by the integral of $\kappa$ over the observed portion of the image plane $\Theta$:
\begin{equation}
M = \Sigma_{cr}D_d^2\int_\Theta \kappa(\vec{\theta})d^2\theta,
\end{equation}
where $D_d$ is the angular diameter distance of the lens relative to the observer.
The surface mass density is subject to the invariance transformation (\ref{invariance}).
The expectation value of the measured mass $\hat{M}$ is
\begin{equation}
{\rm E}\{\hat{M}\} = \Sigma_{cr}D_d^2 {\rm E} \left\{ \int_\Theta \left[\hat{\lambda}\kappa(\vec{\theta}) + w(1 - \hat{\lambda})\right]d^2\theta\right\} = M,
\end{equation}
where we made use of (\ref{nobiassulambda}), and $\Sigma_{cr} \equiv \Sigma_{cr}(z_s \rightarrow \infty)$. {\em If the estimator for $\lambda$ is not biased, then there is also no bias on the estimate of the total mass}.

Let us consider the variance of $\hat{M}$. A straightforward calculation based on Eq. (\ref{nobiassulambda}) gives
\begin{equation}\label{varianzaM}
{\rm Var}(\hat{M}) = {\rm E}\{\hat{M}^2\} - {\rm E}^2\{\hat{M}\} = \left(M - w\Sigma_{cr}D_d^2\int_\Theta d^2\theta\right)^2{\rm Var}(\hat{\lambda}).
\end{equation}
By taking the square root of (\ref{varianzaM}) and dividing it by $M$ we obtain the expected relative dispersion of $\hat{M}$:
\begin{equation}\label{dispersionM2}
\frac{\sigma(\hat{M})}{M} = \left|w\frac{1}{\bar{\kappa}}- 1\right|\sigma(\hat{\lambda}),
\end{equation}
where $\bar{\kappa}$ is the average surface mass density inside the field of view and $\sigma(\hat{\lambda})$ is the square root of (\ref{varianzalambda}), namely the dispersion of $\hat{\lambda}$.

This result allows us to calculate, for a given lens, the accuracy with which its mass can be measured.
For fixed average surface mass density $\bar{\kappa}$, the only quantity that determines this accuracy is the precision in the determination of $\lambda$, $\sigma(\hat{\lambda})$.
This quantity might depend on the shape of the lens mass distribution, so that lenses with certain characteristics may be more suitable candidates than others.
This issue, which is important for determining which are the ideal lens candidates for an application of the present technique, is addressed in the next subsection by studying $\sigma(\hat{\lambda})$ and $\sigma(\hat{M})$ for different lens models.

\subsection{Simple examples}

\begin{figure}[!h]
\resizebox{\hsize}{!}{\includegraphics{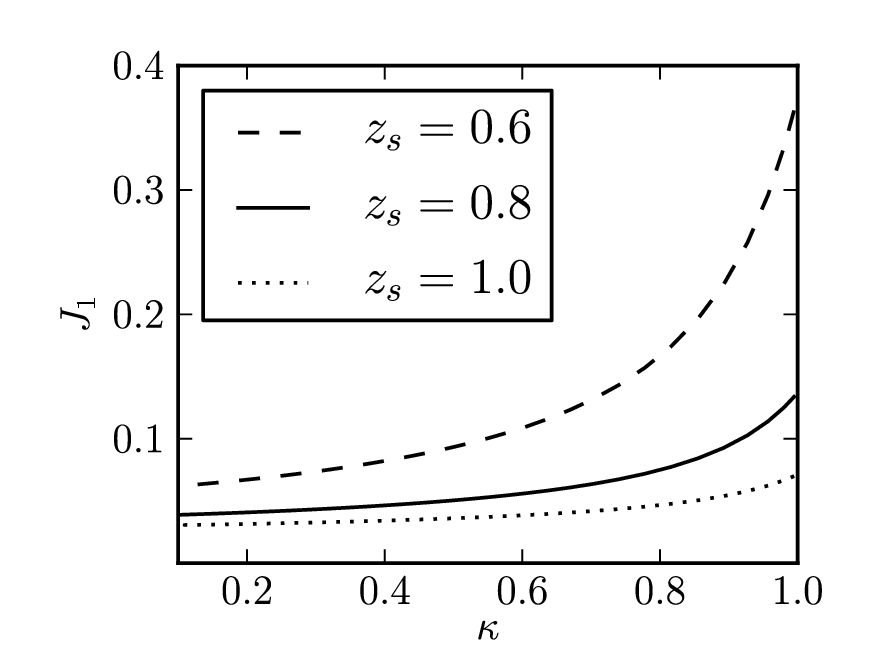}}
\caption{$J_1$, as defined in (\ref{geiuno}), as a function of $\kappa$ (i.e. image position) for a NIS lens with $\sigma_v = 1000 \mbox{ km s}^{-1}$ and $r_c = 58.8 \mbox{ kpc}$, for three source redshifts.}\label{J1NIS}
\end{figure}

To better understand the above results, it is useful to focus on simple situations.
We start by discussing the case with $N_{\mathrm{FP}} = 1$ and consider the quantity in brackets in (\ref{varianzalambda}):
\begin{equation}\label{geiuno}
J_1 \equiv \frac{1}{\displaystyle \frac{1}{\sigma_1^2}[\kappa(\vec{\theta}_1)-w]^2}
\end{equation}
Then, as a first example, we consider an approximate representation of the Nonsingular Isothermal Sphere (NIS), with density profile given by
\begin{equation}\label{NIS}
\rho(r) = \frac{\sigma_v^2}{2\pi G}\frac{1}{r^2 + r_c^2}.
\end{equation}
The projected mass density $\kappa(\vec{\theta})$ of a NIS is given by
\begin{equation}\label{kappaNIS}
\mathbf{\kappa(\vec{\theta}) = \frac{\sigma_v^2}{2G\Sigma_{cr}}\frac{1}{\sqrt{1 + (\theta D_d/r_c)^2}} \equiv \kappa_c\frac{1}{\sqrt{1 + (\theta D_d/r_c)^2}} }
\end{equation}
We take $\sigma_v = 1000 \mbox{ km s}^{-1}$ and $r_c = 58.8 \mbox{ kpc}$ and place the lens at redshift $z_d = 0.3$, so that the projected surface mass density at the center is $\kappa(0) = 1$.
In Fig. \ref{J1NIS} we plot the quantity $J_1$ as a function of $\kappa$ (i. e. image position) for a source at redshift $z_s = 0.6, 0.8, 1.0$. For this and the following examples, the value of $w$ is computed by assuming a redshift distribution of weak lensing measurements $P_z^{(\mathrm{WL})}(z) \propto z^2\exp(-z/z_0)$ with $z_0 = 2/3$, as in Brada\u{c} et al. (\cite{Bradac04}), while the scatter of the Fundamental Plane is assumed to be $\sigma_{\mathrm{FP}} = 0.15$.
It can be seen that $J_1$ increases with increasing $\kappa$, although mildly.
This result suggests that, for a given lens observed within a given field of view, it would be preferable to perform Fundamental Plane measurements on objects whose images lie where the surface mass density is lower, as the expected dispersion on $\lambda$ is lower.
However, it is important to recall the assumptions that underlie this result. In particular, we are assuming that no error comes from the weak lensing analysis.
When this assumption is dropped the situation changes. Real cases are more complex: as will be shown in Sect. 6, weak lensing reconstructions tend to underestimate the surface mass density in the central parts of lenses and overestimate it in the outskirts. 
Therefore, normalizing the overall mass scale of a lens with Fundamental Plane measurements limited to particular regions of the image plane can be risky, because biases can be introduced. Weak lensing errors can be better tackled by sampling the image plane uniformly. This issue is discussed further in Sect. 6.2.

%
%
%
%
%
%
%
%
%
As a second step, we will consider the variation of $\mathrm{Var}(\hat{\lambda})$ as a function of the average surface mass density $\bar{\kappa}$ of the lens, for two simple lens models, again in the case $N_{\mathrm{FP}} = 1$.
As lens models we consider NIS lenses with various values of $\sigma_v$ and $r_c$ but fixed central surface mass density $\kappa_c = 1$.
%
%
%
%
%
%
%
%
%
Equation (\ref{kappaNIS}) shows that NIS lenses with different values of $r_c$ and fixed $\kappa_c$ are rescaled just in the angular dimension.
Therefore, for a given lens, changing $r_c$ and keeping $\kappa_c$ fixed is equivalent to choosing a different field of view.

\begin{figure}[!h]
\resizebox{\hsize}{!}{\includegraphics[width=9cm]{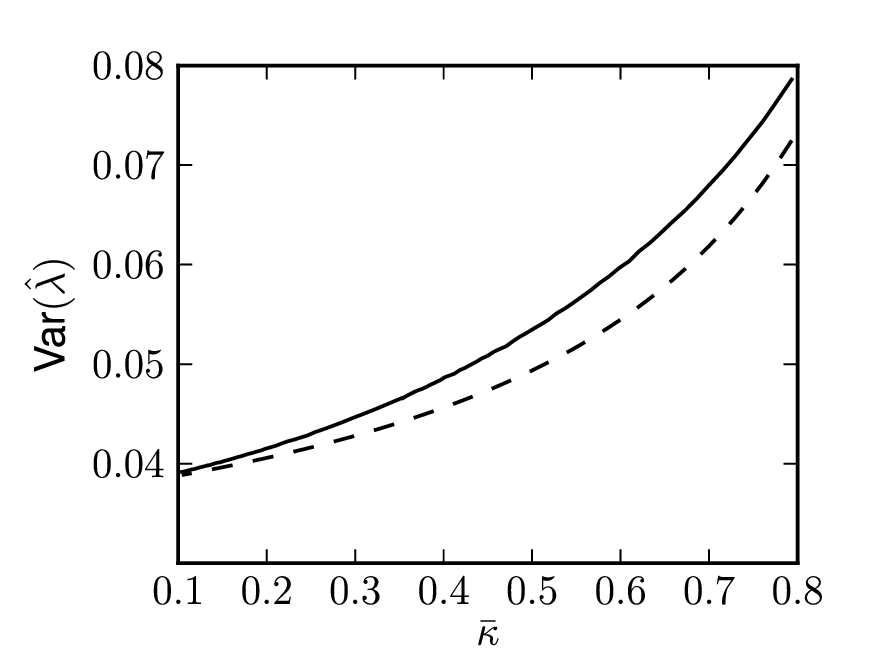}}
\caption{$Var(\hat{\lambda})$ as a function of the average surface mass density $\bar{\kappa}$ for two lens models: a NIS with $\kappa(0) = 1$ (solid line) and a sheet of constant surface mass density and no shear (dotted line).}\label{VarNIScost}
\end{figure}

 %
 %
 %
 %
 %
 %
 %
 %
 %
 For comparison, we also considered lenses of constant surface mass density $\kappa(\vec{\theta}) = \bar{\kappa}$ and no shear.
In Fig. \ref{VarNIScost} we plot the variance of $\hat{\lambda}$ as a function of $\bar{\kappa}$ for these two lens models, for a single source at fixed redshift $z_s = 0.8$. 
The space average in Eq. (\ref{varianzalambda}) was calculated by ignoring effects of the magnification on the image position probability $P_{\vec{\theta}}(\vec{\theta})$.
The value of the variance of the estimator $\hat{\lambda}$ is found to be similar for the two kinds of lens, for fixed average density $\bar{\kappa}$.
This result is presumably a consequence of the mild dependence of $J_1$ on $\kappa$, as observed in the plot of Fig. \ref{J1NIS}.

\begin{figure}[!h]
\resizebox{\hsize}{!}{\includegraphics{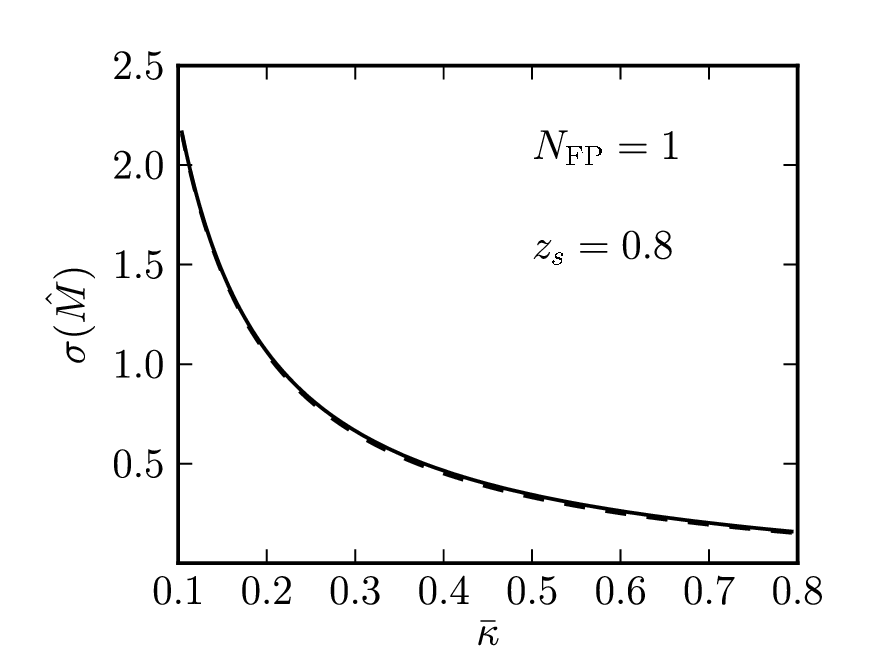}}
\caption{$\sigma(\hat{M})/M$ as a function of the average surface mass density $\bar{\kappa}$ for two lens models: a NIS with $\kappa(0) = 1$ (solid line) and a sheet of constant surface mass density and no shear (dotted line).}\label{sigmaM}
\end{figure}

Finally, we calculate the dispersion on the measurement of the mass, $\sigma(\hat{M})/M$ for the two cases considered above, plotting the results in Fig. \ref{sigmaM}. 
The results for the two lenses are practically indistinguishable.
This result, which reflects the similarity of the values of $\sigma(\hat{\lambda})$ found for the two lenses, suggests that the shape of the mass distribution plays little role in determining the accuracy of the mass measurement, while the decisive factor is the average surface mass density within the field of view, $\bar{\kappa}$.

We recall that these results were obtained by assuming that the image positions are distributed randomly.
This may not be true if the observer plays an active role in the object selection. For example one may wish to sample the field uniformly, avoiding close pairs of images. In this way the image positions will be correlated. This will affect the value of $\sigma(\hat{\lambda})$.
However, given the already mild dependence of $\mbox{Var}(\hat{\lambda})$ on the image position (see Fig. \ref{J1NIS}), we do not expect our results to change significantly.

\section{Substructure effects}\label{substructures}

In some galaxy-galaxy strong lensing systems it has been noted that while it is relatively easy to build smooth lens models that reproduce well the multiple image positions of distant QSOs, the same models are unable to reproduce the observed flux ratios of these images (e.g. Kent \& Falco \cite{KentFalco}, Hogg \& Blandford \cite{Hogg}, Falco et al. \cite{Falco97}). It is now common belief that these flux ratio anomalies may be due to the presence of substructures, and several attempts have been made at providing suitable lens models that take substructure into account, such as those by Mao \& Schneider (\cite{MaoSchneider}), Metcalf \& Madau (\cite{MetcalfMadau}), Brada\u{c} et al. (\cite{Bradac02}), Chiba (\cite{Chiba}).
Their argument is supported by the following reasoning: a substructure having an Einstein radius comparable to the angular size of the source's light emitting region can cause a significant change in its apparent size, while its position on the lens plane can be little affected. 
This condition is relatively easy to obtain for the case of a compact source such as a distant QSO being lensed by a galaxy; in fact, objects as massive as a typical globular cluster ($\sim 10^6 M_\odot$) can change appreciably the flux received from such a source, if properly aligned.
This effect is enhanced in the proximity of critical curves (i.e. curves where $|\det A| \simeq 0$).
Since
\begin{equation}
\mu = |\det A|^{-1} = \left|\frac{1}{(1-\kappa)^2 - |\gamma|^2}\right|,
\end{equation}
if $|\det A| \simeq 0$ a small change in $\kappa$ can produce a large change in the magnification.

Weak lensing is a good tracer of the surface mass density averaged over finite portions of the image plane, but is not sensitive to small scale variations of the projected mass distribution.
In fact, weak lensing methods recover the reduced shear $g$ by averaging the distortion signal over a number of background galaxies over angular scales of tens of arcseconds (e.g. see Lombardi et al. \cite{Lombardi2000}). Therefore, they only provide smoothed mass density profiles.
If we want to break the mass-sheet degeneracy with magnification measurements, we must be sure that these magnification measurements also reflect the properties of this smoothed mass profile.
Substructures modify the lensing signal in a nontrivial way and can complicate the interpretation of magnification measurements.
For the purpose of constraining the total mass of the lens, the contribution from smaller clumps has the same effect of noise.

The problem is addressed with the aid of numerical simulations.
We construct a model cluster as a superposition of a smooth principal halo and a number of smaller subhalos. 
We generate images of background early--type galaxies, compare the observed magnification with that inferred from a smooth lens model, and then analyze the results.

In this section we will use the terms clump, halo, subclump, subhalo, substructure synonymously to refer to mass concentrations inside a galaxy cluster.

\subsection{Modelling substructure in clusters of galaxies}\label{ModelSubstructure}

It is not clear how much substructure is present in clusters of galaxies. 
Cosmological simulations predict the existence of a large number of subhalos of mass $M < 10^{10} M_\odot$, but observations have failed to prove their presence, so far.
Here we will adopt a conservative approach and will take into account the possibility that substructures are present in the abundances predicted by $\Lambda$CDM models.

\subsubsection{Masses and spatial distribution of subclumps}

N-body simulations of structure formation at cluster scales based on the $\Lambda$CDM scenario have shown that the dark matter halos that maintain their identity after the formation process can be approximately described by a power-law mass function:
\begin{equation}\label{MassFunction}
\frac{dN}{dM} \propto M^{-(1+\alpha)}
\end{equation}
with $\alpha \approx 0.9\div 1$, as demonstrated for example in the papers by Tormen et al. (\cite{Tormen}), Ghigna et al. (\cite{Ghigna}), De Lucia et al. (\cite{DeLucia2004}) and Gao et al. (\cite{Gao}). 
These dark matter subclumps are believed to account for a few percent of the total mass of the cluster. Note that the mass fraction in substructure also depends on the age of the cluster: as a cluster evolves the subhalos merge to the main halo until the whole structure virializes and fewer substructures are left.

Given these results, we will model clusters in a semi-analytic way.
A smooth mass density profile for the main halo is chosen;
a distribution of subclumps is randomly generated from the mass distribution given by (\ref{MassFunction}), with the total mass in substructures fixed;
these subclumps are randomly distributed with a spatial probability distribution proportional to the mass density of the main halo.
More details about the practical realization of this procedure are given below.

Numerical simulations have also shown that more massive clumps tend to be located far from the cluster center, where only small scale halos survive the merging process (Tormen et al. \cite{Tormen}, Ghigna et al. \cite{Ghigna}). 
For simplicity we adopted a uniform spatial distribution of subhalos, as appropriate for the kind of study we wish to perform.

\subsubsection{Internal structure of subclumps}

For the description of the internal structure of dark matter subhalos we follow the work of Metcalf \& Madau (\cite{MetcalfMadau}), who developed numerical simulations to study the effects of subclumps in galaxies on the measured lensing magnification of distant QSOs.
In our work we extend the use of their tools to galaxy clusters environments.
Metcalf \& Madau modeled subclumps as truncated singular isothermal spheres (TSIS). The advantage of using singular isothermal spheres is that their lensing properties can be easily described analytically.
On the other hand a singular isothermal sphere has infinite mass. For this reason a truncation radius is introduced.
For a clump of mass $m$ at radial distance $R$ from the center of the main halo the truncation radius is taken to be equal to the {\em tidal radius} $r_t$, which is estimated as (Metcalf \& Madau \cite{MetcalfMadau})
\begin{equation}\label{tidalradius}
r_t \simeq R\left[\frac{m}{3M(R)}\right]^{1/3},
\end{equation}
where $M(R)$ is the mass of the main halo enclosed by the sphere of radius $R$. 
With this choice, if $M(R)$ grows less steeply than $R^3$ (both SIS and NFW models have this property) then clumps closer to the center tend to be more compact than those that lie far from the center.

Once the mass and truncation radius of the clump are fixed there is a unique truncated singular isothermal sphere with those characteristics. In particular, if we adopt the following notation for the density of a TSIS,
\begin{equation}
\rho(r) = \left\{\begin{tabular}{ll}$\displaystyle\frac{\sigma_v^2}{2\pi G r^2}$ & if $r < r_t$ \\
 & \\
0 & elsewhere
\end{tabular}\right .,
\end{equation}
the parameter $\sigma_v$ (often called the velocity dispersion) is given in terms of the mass $m$ of the clump by
\begin{equation}
\sigma_v^2 = \frac{Gm}{2r_t} = \frac{G m^{2/3}}{2R}\left[3M(R)\right]^{1/3}.
\end{equation}

\subsection{Lensing simulations}\label{4Simulations}

\subsubsection{General prescriptions}

A simulated cluster is generated with the procedure described in Sect.~\ref{ModelSubstructure}. For the main halo we adopt a nonsingular isothermal sphere (\ref{NIS}).
Subhalos with a mass distribution given by (\ref{MassFunction}) and cumulative mass that accounts for a fraction $f$ of the total mass of the cluster within $r_{200}$ are then added to this smooth component.
Here $r_{200}$ represents the radius of the sphere whose mean density is 200 times the critical density at the redshift of the object, and $M_{200}$ is the corresponding mass. The total mass in subhalos is then $fM_{200}$. 
A direct use of (\ref{MassFunction}) results in a very large number of clumps with very small mass. These small clumps are not relevant for the lensing problem because they produce negligible effects on the magnification of extended images such as those of early-type galaxies. Moreover, the inclusion of a large number of clumps increases the computational effort required to run the simulations, so that it is useful to introduce a cutoff on the lower range of possible masses. A reasonable choice for this lower mass cutoff is a value $m_{min}$ for which the typical lensing deflection angle is only a small fraction of the angular size of the source considered.

%
%
%
%
%
%
%
%
%
%
%
%
%
%
%
%
%
%
To test how the results of our simulations depend on the assumed internal structure of clumps we also adopted an alternative (and non realistic) internal profile: we assumed clumps to be point masses.
Clearly, the TSIS clump will produce a smaller deflection angle up to $r_t$. For distances larger than $r_t$ the two clumps will produce equal values of the deflection angle, as they both act as if they were point masses located at the center.
For this reason, the effects on magnification produced by a point mass will be generally higher than those produced by an extended lens of the same mass.
Therefore, if we study the problem of the magnification induced by substructure by modelling substructure as point masses we can obtain an upper limit on the effects of substructure on the magnification.

\subsubsection{Practical realization}

We simulate the observed image of a circular early-type galaxy with effective radius $R_e = 5\,{\rm kpc}$ at redshift $z_s = 0.8$ being lensed by a NIS cluster with substructures at redshift $z_d = 0.3$.
The parameters of the NIS model chosen for the test of the method are taken from a study of the Coma cluster by De Boni \& Bertin (\cite{DeBoni}), who found the following best fit parameters for the description of the dark matter halo as a NIS model: $r_c = 88\,{\rm kpc}$ and $\sigma_v = 1156 \mbox{ km s}^{-1}$.

We fix the mass fraction in substructure $f$ and generate substructures with the procedure described above, with a lower mass cutoff of $m_{min} = 10^9 M_\odot$. This value was chosen because, for the redshift configuration of our system, the Einstein radii of point masses less massive than $m_{min}$ are smaller than $0.3\,{\rm kpc}$, thus with little impact on the distortion of the images chosen for our study.
Note that in a situation in which all the substructures account for a fraction $f$ of the total mass of the cluster, those with mass greater than $m_{min}$ will in general account for a lower fraction $f'$ of the total mass. A fraction $f - f'$ will consist of clumps with $m < m_{min}$. 
Since the lensing effect of these low mass mass clumps is small we simulate them by adding a smooth component for a fraction $f - f'$ of the total mass.
We generated clusters with values of $f$ equal to 0.05, 0.10 and 0.15. In these three situations and with the adopted value of $m_{min}$ the corresponding value of $f'$ is 0.029, 0.061 and 0.093 respectively.

To simulate the image formation we proceed as follows, taking inspiration from the work by Metcalf \& Madau (\cite{MetcalfMadau}).
We define a field of view of $4\times4\,\mathrm{arcmin}^2$ centered on the main halo. We then define a grid over the entire field of view. The resolution of the grid is such that the observed effective radius for the early-type galaxy sources in the absence of magnification is 10 grid cells long.

At every grid point the lensing deflection angle is computed. 
The contribution of the subclumps is calculated in the following way:
the clumps are first assumed to be point masses and the deflection angle generated by each clump is calculated at every grid point. Then, only for the clumps that lie within the field of view considered, the deflection angle inside the circle of radius $r_t$ is corrected with the expression for the deflection angle of a TSIS lens (Metcalf \& Madau \cite{MetcalfMadau}):
\begin{equation}\label{alphaTSIS}
\alpha(x) = \alpha_0\left\{\begin{tabular}{ll} $\displaystyle \frac{1}{a} - \sqrt{\frac{1}{a^2}-1} + \arctan{\sqrt{\frac{1}{a^2}-1}}$ & if $a < 1$ \\
& \\
$\displaystyle \frac{1}{a}$ & if $a > 1$\end{tabular} \right .
\end{equation}
where $a \equiv x r_c/r_t$, $x \equiv r/r_c$. More simply, $a = r/r_t$.
Differently from the work of Metcalf \& Madau (\cite{MetcalfMadau}), we calculate the deflection angle once and for all the grid points in the field, taking into account the contribution to $\vec{\alpha}$ of every subhalo in the lens.

Once the map of the deflection angle over the field of view is created, images of early-type galaxies are generated with a uniform distribution on the lens plane. 
%
%
%
%
%
%
%
%
%


Effects of the magnification on the image position distribution are neglected, i. e. a uniform distribution in the image plane is adopted.
A rigorous treatment of this aspect would require knowing the luminosity function of our target galaxies, but this is beyond the goals of this paper. 
Nevertheless, we explored different scenarios by applying importance sampling to the outcome of our simulations.
In particular, we examined two different probability distributions: proportional to the magnification, and inversely proportional to the magnification. 
The change in the results with respect to the uniform case was less than 15\%, indicating that our approximation does not alter the conclusions of our study.




%
%
%
%
%
%
%
%
%
Sources are described with circular de Vaucouleurs profiles, truncated at $r_e$. Images are then constructed in the following way.
We define a square in the image plane centered on the adopted location for the image center. The size of this square is chosen so that the length of its sides are a few times the effective radius times the linear stretching ($\sim \mu^{1/2}$) produced by the smooth component of the lens. In this way the observed image is guaranteed to lie within this square.
Then, each pixel inside the square is mapped to the source plane with the lens equation of the simulated cluster, where the deflection angle was calculated with the procedure described above. 
Since gravitational lensing preserves surface brightness, the brightness of each pixel in the image plane is taken to be equal to the brightness of the corresponding point of the source plane to which it is mapped. 
Pixel mapped outside the circle of radius $r_e$ around the center of the source do not belong to the observed image.
The observed magnification is then defined as the ratio between the total brightness of the image and that of the source. This gives us the flux magnification, which is equal to the size magnification because of the conservation of surface brightness.

This procedure introduces an error related to the pixelization of the definition of the observed image. The relevance of this error source was checked while running our simulations.
In particular we first performed the simulations for a smooth lens without substructures, and compared the magnification measured with the above procedure with the theoretical value of the magnification in the image centroid, given by the model. The results of this test are the following:
\begin{equation}
\frac{\Delta\mu}{\mu} = 0.018\,\, \qquad \frac{\sigma(\mu)}{\mu} = 0.027,
\end{equation}
where $\Delta\mu/\mu$ and $\sigma(\mu)/\mu$ are the average relative offset and standard deviation between the theoretical and measured magnifications, respectively.
The test was performed with a grid resolution such that the length of a pixel is a tenth of $r_e$.
The resulting typical error in the magnification is of order of a few percent. This is much smaller than the $\sim 30\%$ error expected for a measurement of the magnification of an early-type galaxy with the use of the Fundamental Plane relation. Given this result, we adopted the same grid resolution for the actual simulations.

In the simulations with substructures, a small fraction of the images display strong lensing features (multiple images, arcs, or rings). This is made possible by the fact that the deflection angle for a TSIS lens approaches a finite value as the distance to the center approaches zero, as can be seen from (\ref{alphaTSIS}).
For such strongly lensed images the observed magnification calculated with the above method is very different from the magnification obtained with a smooth lens model. Such events are rather rare (typically a few per 1000 images).
On the other hand, if such situation occurred in an actual observation it would be easily recognized as a strong lensing feature.
A highly distorted image, as an arc is, would not be suitable for a Fundamental Plane measurement. Therefore {\em these images are removed from our analysis}.

At the same time we also took care not to make an excessive use of this procedure, for the following reason.
The capability of a clump of a given mass to form arcs and multiple images depends on its internal structure.
In our model we assumed TSIS mass profiles, but the real case is likely to be different from that.
In fact, Meneghetti et al. (\cite{Meneghetti2003}) showed that semi-analytic models with NFW subhalos provide a better agreement with the observed statistics of arcs than models with SIS subclumps.
Then, if in our simulation we observe an arc created by the presence of a massive subhalo it could be that the same clump with a more realistic internal structure would have not produced an arc but only a highly distorted image whose magnification could have been measured.
For such images the value of the observed magnification is likely to differ significantly from the value inferred by assuming a smooth mass distribution.
Thus by eliminating them from the analysis we would bias the results towards a better accordance between observed magnifications and smooth model magnifications.
This problem is more relevant in the simulations in which clumps are treated as point masses, as they are more capable of producing arcs.

On the other hand the most significant departures of the observed magnifications from the smooth case are for images in the proximity of the most massive ($m > 10^{11} M_\odot$) subclumps. 
This is in agreement with the work by Meneghetti et al. (\cite{Meneghetti}, see discussion in 6.2.1).
These are galaxy-scale objects and it is unlikely that they exist only in dark form.
In other words, such substructures are likely to be easily identified by the presence of a luminous component. 
Therefore, for an image that lies in the proximity of one such object we should immediately suspect that part of the observed magnification is caused by the presence of this substructure and we would be warned of the bias caused by interpreting this data as tracer of the smooth component of the cluster. Moreover, a number of gravitational lensing studies have been presented in which the lensing contribution of individual galaxies was incorporated into the analysis (see e.g. Natarajan et al. \cite{Natarajan2004}). Therefore, the same thing could be done for images of early-type galaxies close to cluster member galaxies, which is the situation discussed here. 

%
%
%
%
%
%
%
%
%
When multiple images are observed (such as the pair in Fig. \ref{multipleimage}), we take the magnification measurement obtained with the larger image only.
In a realistic situation, if Fundamental Plane measurements are carried out on both the images, it would be relatively easy to label them as a pair and then to infer the presence of a substructure.
We decided to adopt the more conservative approach in which the observer does not see the counterimage.


%
%
%
%
%
%
%
%
%
\begin{figure}[!h]
\begin{center}
\resizebox{\hsize}{!}{\includegraphics[width = 8cm]{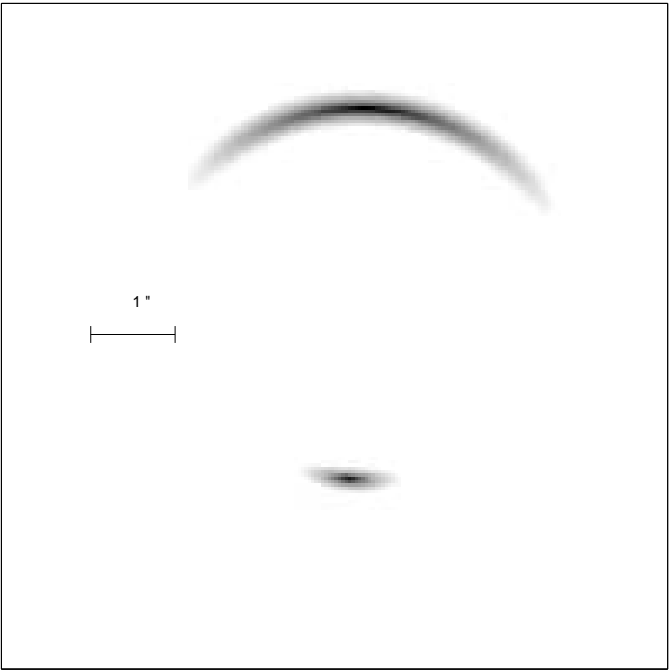}}
\caption{Strongly lensed image in the proximity of a massive subhalo. Such images are removed from our analysis. This is done only after making sure that the arc feature would be recognizable under realistic observing conditions, i.e. that the arc-like shape would not be smeared out by a realistic PSF. In this example the larger arc is a few arcseconds long and could be easily identified.}\label{arcimage}
\end{center}
\end{figure}

\begin{figure}[!h]
\begin{center}
\resizebox{\hsize}{!}{\includegraphics{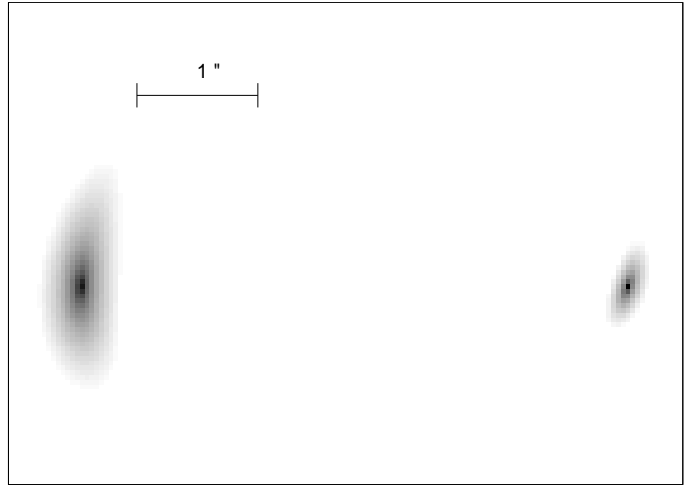}}
\caption{Multiple image system. This image (and others with similar properties) is not discarded, since the shape of the image on the left is not so strongly distorted and in general it may not be recognized as part of a multiple image system. In such cases only the larger image is considered for the analysis.}\label{multipleimage}
\end{center}
\end{figure}

\subsection{Results}

We studied image magnifications with three realizations of simulated clusters, with mass fraction in substructure $f$ of 0.05, 0.10 and 0.15. For each case we adopted two different models for the internal structures of the subclumps: TSIS and point mass.
For each case we generated $N_s = 10000$ sources with the procedure described above.
For each source, we studied the observed magnification with the presence of substructure, $\mu_{subs}$, and the magnification that would be observed with a smooth mass distribution, $\mu_{smooth}$. 
In Tables \ref{risultatisimsub1} and \ref{risultatisimsub2} we report the differences between the measured values of $\mu_{subs}$ and $\mu_{smooth}$.
In particular, we introduce the average relative difference, defined as
\begin{equation}
\frac{\Delta \mu}{\mu} \equiv \frac{1}{N_s}\sum_{i=1}^{N_s} \frac{\mu_{subs,i} - \mu_{smooth,i}}{\mu_{smooth,i}},
\end{equation}
and the expected mean relative error,
\begin{equation}
\frac{\sigma_{\mu}}{\mu} \equiv \sqrt{\frac{1}{N_s}\sum_{i=1}^{N_s}\left(\frac{\mu_{subs,i} - \mu_{smooth,i}}{\mu_{smooth,i}}\right)^2}.
\end{equation}
For completeness we also record the number $N_{rej}$ of rejected images in each simulation run.

\begin{table}
\caption{TSISs subhaloes}
\label{risultatisimsub1}
\begin{tabular}{c|c c c}
$f$ & ${\Delta\mu}/{\mu}$ & ${\sigma_{\mu}}/{\mu}$ & $N_{rej}$\\
\hline
0.05 & 0.005 & 0.061 & 4 \\

0.10 & 0.014 & 0.077 & 6\\

0.15 & 0.018 & 0.091 & 20\\
\end{tabular}
\
\tablefoot{Diffefrences between magnifications in the presence and in the absence of substructure, for three cluster realizations. Clumps are modeled as TSISs. Each simulation run was performed by generating 10000 images. $N_{rej}$ is the number of strongly lensed images that were rejected in each run.}
\end{table}

\begin{table}
\caption{Point mass subhaloes}
\label{risultatisimsub2}

\begin{tabular}{c|c c c}

$f$ & ${\Delta\mu}/{\mu}$ & ${\sigma_{\mu}}/{\mu}$ & $N_{rej}$\\
\hline
0.05 & -0.014 & 0.052 & 41 \\

0.10 & 0.002 & 0.063 & 104\\

0.15 & 0.027 & 0.075 & 127\\
\end{tabular}
\tablefoot{Differences between magnifications in the presence and in the absence of substructure, for three cluster realizations.  Clumps are modeled as point masses. Each simulation run was performed by generating 10000 images. $N_{rej}$ is the number of strongly lensed images that were rejected in each run.}
\end{table}

The quantity that is most relevant for our study is the dispersion of observed magnification around the expected value, $\sigma_{\mu}$. As expected, this quantity increases with increasing mass fraction in substructure. However, the values of this dispersion are somehow small.
This is rather good news, because it means that the magnification of early-type galaxies is more sensitive to the smooth component of the mass distribution, which accounts for the bulk of the mass, and is therefore a quantity suited to constrain the total mass of the lens.
%
%
%
%
%
%
%
%
%
%
%
%
%
%
%
%
%
%

The case of point mass substructures deserves further discussion.
As noted above, point masses are expected to produce larger differences between the observed magnification and $\mu_{smooth}$. 
From a first look at the results of Tables \ref{risultatisimsub1} and \ref{risultatisimsub2}, it seems that the opposite situation is realized.
However, in the simulation with point mass substructures the number of strongly lensed images that is rejected is higher than in the TSIS case.
This means that a significant fraction of the sources whose images are rejected in the point mass case would be included in the analysis if lensed by a model with TSIS substructures.
Presumably, the images of these sources are magnified by substructures and they contribute significantly to the value of the measured dispersion.
This means that in the TSIS case the observed higher value of the dispersion is determined by a small number (less than 1 in 100) of images, and by excluding them from the analysis we would obtain a dispersion not larger than the one observed in the point mass simulation.
This result is in qualitative agreement with the work by Meneghetti et al. (\cite{Meneghetti}). They showed that in a more realistic cluster realization the probability of having a tangential-to-radial magnification (equivalent to observed axis ratio for circular sources) larger than 5 is less than 3\% (see Fig. 7 of the cited paper). They also claim that, in their case, substructures account for 30\% of the strong lensing cross section (the smoothed version of their cluster is still a critical cluster, unlike our case), meaning that the above percentage must be scaled accordingly to be compared with our study.

On the basis of these results we conclude that the magnification of early-type galaxies is little influenced by the presence of substructure. Substructure seems to play a significant role in the image formation only when present in large amounts, which is an unlikely scenario.
Current estimates of the mass fraction in substructure based on numerical simulations give as typical values $f \lesssim 0.10$ (Tormen et al. \cite{Tormen}; Ghigna et al. \cite{Ghigna}).
These results give more significance to the technique of magnification measurement based on the use of the Fundamental Plane relation, and set a solid base for the adoption of the technique described in this paper for the purpose of solving the problem of the mass-sheet degeneracy.

\section{Testing the method}\label{Simulations}

The goal of the study presented in this section is to clarify to what extent the results obtained in Sect.~\ref{statistics} also hold in more realistic situations.

We take a model cluster lens, apply the mass measurement technique to synthetic weak lensing and magnification data simulated for this model, and then we analyze the results obtained.
In particular, we compare the dispersion on the measurement of the mass obtained in these simulations with the dispersion expected by considering the effects of Fundamental Plane measurements only, obtained from (\ref{dispersionM2}): if the two quantities will not differ substantially, then it means that weak lensing errors do not play an important role and that the theoretical treatment of Sect.~\ref{statistics} can find applications in practical cases.

Before facing the problem in full it is interesting to study how the errors in the weak lensing analysis alone influence the estimates of the total mass of the lens.
In other words, we wish to clarify what is the typical error in the estimate of $M$ in the hypothetical case of perfect Fundamental Plane measurements.
This situation is simulated first and a realistic case in its full aspects is studied later.

\subsection{Simulations}

In the following we will describe the ingredients necessary to set up the simulations.

\subsubsection{The lens model}

The model adopted to describe the lens is a Nonsingular Isothermal Sphere (NIS).
The choice of a smooth model for the lens is suggested by the results of the analysis described in Sect.~\ref{substructures} on the effects of substructures on the magnification of the images of early-type galaxies.

The parameters of the NIS model chosen for this test are the same adopted for the description of the smooth component of the lens model used in the simulations of Sect.~\ref{substructures}: $r_c = 88\,{\rm kpc}$ and $\sigma_v = 1156 \mbox{ km s}^{-1}$.

\subsubsection{Weak lensing data}

For the simulation of weak lensing measurements we take inspiration from the paper by Brada\u{c} et al. (\cite{Bradac04}).
The procedure adopted is the following.
\begin{itemize}
\item We generate background galaxies with a uniform spatial distribution in a $4\times 4\,\mathrm{arcmin}^2$ field of view (magnification effects on the spatial distribution of images are neglected). Three different values of the number density are chosen: $n = 30,\,50,\,70\,\mathrm{arcmin}^{-2}$.
\item Each galaxy is assigned a redshift, taken from the following distribution:
\begin{equation}
P_z^{(\mathrm{WL})}(z) \propto z^2 e^{-z/z_0},
\end{equation}
with $z_0 = 2/3$, as suggested by Brainerd et al. (\cite{Brainerd}). 
This is a standard choice for weak lensing simulations.
\item Each background galaxy is then assigned an intrinsic ellipticity drawn from a truncated Gaussian distribution:
\begin{equation}
P_{\epsilon^s}(\epsilon^s) = \frac{1}{2\pi\sigma_\epsilon^2[1 - \exp{(-1/2\sigma_\epsilon^2)}]}\exp{\{-|\epsilon^s|^2/2\sigma_\epsilon^2\}},
\end{equation}
with $\sigma_\epsilon = 0.25$. This is also a standard choice for weak lensing simulations (e.g. see Bartelmann et al. \cite{Bartelmann96}, Seitz \& Schneider \cite{SS97}).
\item For each galaxy, the resulting ellipticity is calculated from Eq. (\ref{ellipticity}).
\item An artificial measurement error is added, so that the resulting measured ellipticity $\epsilon^m$ is
\begin{equation}
\epsilon^m = \epsilon + \epsilon^{err},
\end{equation}
where $\epsilon^{err}$ is a random error generated from a Gaussian distribution with dispersion $\sigma_{err} = 0.1$. In adding the errors we ensured that $|\epsilon^m| < 1$.
\end{itemize}
The data are then processed with the finite-field inversion technique of Seitz \& Schneider (\cite{SS97}).
A grid in the image plane is defined.
At each grid point $\{i,j\}$, the average ellipticity $\left<\epsilon\right>$ is estimated from the observed ellipticities $\epsilon_k$ of background galaxies as
\begin{equation}
\left<\epsilon\right>_{i,j} = \sum_{k=1}^{N_g} \epsilon_kW(|\vec{\theta}_{i,j} - \vec{\theta}_k|),
\end{equation}
where $W(\vec{\theta})$ is a Gaussian with dispersion $\Delta\vec{\theta}$, such that $n\Delta\vec{\theta}^2 = 12$.

\subsubsection{Fundamental Plane measurements}

After the weak lensing reconstruction, which provides a model density map $\kappa_0(\vec{\theta})$ up to the invariance transformation (\ref{invariance}), the simulation proceeds with the generation of $N_{\mathrm{FP}} = 20$ early-type galaxies and the related Fundamental Plane measurements.
This is done as follows.

The position of each galaxy is generated randomly with a uniform distribution on the image plane (again, the effects of magnification in the spatial probability distribution of images are not considered).
Each galaxy is then assigned a redshift $0.5 < z < 1.0$. The upper limit reflects the redshift limit reached by the Fundamental Plane measurements carried out so far. The lower limit instead is set because the lensing signal for sources too close to the lens is too low.
The simulation also requires a specification of the shape of the redshift distribution of the observed early-type galaxies, $P_z^{(\mathrm{FP})}(z)$.
This quantity depends on the intrinsic luminosity function of early-type galaxies, which in general varies with redshift, and also on the object selection procedure. Given these uncertainties, for our simulations we adopted a uniform distribution to reduce the computational effort.
It is also assumed that no error is introduced in the determination of the individual redshifts.

For each early-type galaxy, the quantity $\sqrt{1/\mu} = r_e^{\mathrm{(FP)}}/r_e^{\mathrm{(obs)}}$ is then generated from a Gaussian distribution with 15\% dispersion, centered on the true value given by the model.
This dispersion should reflect the observed scatter in $r_e$ of the Fundamental Plane relation. In our case we adopted an optimistic estimate of this latter quantity.

\subsection{Results}

Under the observation conditions described above
($N_{\mbox{FP}} = 20$)
, the expected dispersion on the estimate of the total mass of the lens 
%
%
%
%
%
%
%
%
%
%
%
%
%
%
%
%
%
%
, calculated from (\ref{dispersionM2}) and therefore ignoring weak lensing errors, is
\begin{equation}\label{stimaerrore}
\sigma_{\mathrm{exp}} = \frac{\sigma(\hat{M})}{M} =  0.21\left(\frac{\sigma_{\mathrm{FP}}}{0.15}\right),
\end{equation}
where $\sigma_{\mathrm{FP}}$ is the dispersion in $r_e$ of the Fundamental Plane, averaged over redshift.

To better quantify the effects of weak lensing, the simulations have first been performed in the hypothetical case of perfect magnification measurements ($\kappa_i^{(FP)} = \kappa(\vec{\theta}_i)$), from which we obtained an estimate of the dispersion in the measurement of the total mass introduced by weak lensing errors only. The results of the simulations relative to this particular case are presented in \ref{soloweaklensing}, while in \ref{realisticcase} we report those obtained in a more realistic situation, in which the simulated magnification measurements have a 30\% dispersion.

\begin{figure}[!h]
\resizebox{\hsize}{!}{\includegraphics{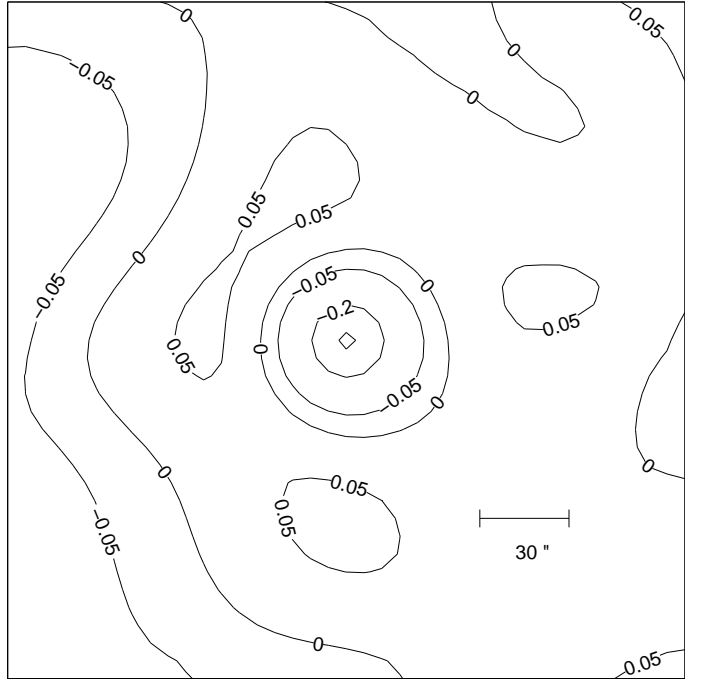}}
\caption{Contour plot of the difference $\kappa_{WL}(\vec{\theta}) - \kappa(\vec{\theta})$, where $\kappa(\vec{\theta})$ is the true surface mass density of the lens and $\kappa_{WL}(\vec{\theta})$ is a weak lensing reconstruction of the lens yielding the same total mass. In the central region the reconstructed profile underestimates the surface mass density, while in a significant region of the image plane the opposite case occurs.}\label{contourdiff}
\end{figure}

Let us consider a mass density map $\kappa_{WL}(\vec{\theta})$ obtained from weak lensing and transformed with (\ref{invariance}) to reproduce the exact value of the lens mass within the field of view.
Since weak lensing reconstructions typically produce smoothed density maps, we expect $\kappa_{WL}(\vec{\theta})$ to underestimate the surface mass density in the central parts of the lens, and to be higher than the true value in the outer parts.
This is indeed what is observed in the simulations: in Fig. \ref{contourdiff} we plot the difference $\kappa_{WL}(\vec{\theta}) - \kappa(\vec{\theta})$ for an example case of weak lensing reconstruction.

This property has some consequences on the process of fitting the density map to Fundamental Plane measurements.
Since the expectation value of $\kappa^{\mathrm{(FP)}}$ is approximately the true surface mass density, measurements close to the center will tend to give estimates $\kappa^{\mathrm{(FP)}}$ higher than $\kappa_{\mathrm{WL}}(\vec{\theta})$ and therefore will bias the measurement of the total mass towards higher values (remember that $\kappa_{\mathrm{WL}}(\vec{\theta})$ was defined as the density map that corresponds to the correct value of the mass). In contrast, Fundamental Plane measurements in parts of the lens plane where $\kappa_{WL} - \kappa > 0$ will tend to bias the mass towards lower values.
Thus, if we distribute the sources uniformly we expect that this potential source of bias can be overcome statistically.
This effect is more evident when there is no error on $\kappa^{\mathrm{(FP)}}$: in that case, the errors in the estimate of the total mass are only due to errors in the weak lensing analysis.

\subsubsection{Ideal case: perfect magnification measurements}\label{soloweaklensing}

1000 simulation runs are performed for three different values of the number density of background galaxies: $n = 30,\,50,\,70\,\mathrm{arcmin}^{-2}$, and two different numbers of Fundamental Plane measurements: $N_{\mathrm{FP}} = 1$ and $N_{\mathrm{FP}} = 20$.

In Tables \ref{soloweaktable1} and \ref{soloweaktable2} the relative mean errors $\Delta M/M = (\bar{M} - M)/M$ and dispersion $\sigma(M)/M$ obtained in these simulations are reported.
As expected, the error in the determination of the total mass decreases as the number density of background galaxies is increased.
However, the dependence of $\sigma(M)$ on $n$ is mild: this means that the number density of background galaxies is not a critical factor in the effectiveness of the method.
It can also be seen that with a single magnification measurement the total mass is poorly constrained, as expected.
This result shows that with a single local estimate of the surface mass density, no matter how accurate, it is difficult to break the mass--sheet degeneracy.

On the other hand, in the case of $N_{\mathrm{FP}} = 20$ Fundamental Plane measurements the dispersion is only a few percent.

Another significant result is that there is practically no bias in the estimate of the total mass ($\Delta M/M \simeq 0$).
This result is clearly a consequence of the assumption of a uniform spatial distribution for the Fundamental Plane measurements.
If we manage to pick a sufficient number of early-type galaxies more or less uniformly distributed in the field of view there are good chances for the final measurement of the mass to be unbiased.
This is a great advantage of the present technique with respect to the use of strong lensing information.
Strong lensing features are typically limited to the central regions of clusters. Then, since the surface mass density in the central parts of a cluster obtained from weak lensing is typically underestimated, the inclusion of strong lensing data might lead to higher estimates of the total mass.
With the present method this effect can be kept under control.
%
%
%
%
%
%
%
%
%

In principle, substructures can introduce noise in the weak lensing signal as well.
The nonparametric reconstruction method used here can recover features of the scale of the smoothing length or larger, as shown by Seitz \& Schneider (\cite{SS97}).
However, King et al. (\cite{King}) and Clowe et al. (\cite{Clowe}) showed that the effects of small scale substructures are of modest importance for weak lensing measurements.
Therefore we can conclude that the results presented here do not depend on our choice of a simplified cluster model.

Additional noise in the weak lensing measurements can be introduced by the presence of uncorrelated large scale structure along the line of sight.
This issue was studied extensively by Hoekstra (\cite{Hoekstra}).
He found that the contribution of structures non associated with the cluster is important at large radii, but is negligible for the relatively small fields of view considered in our work (a few arcminutes).

%
%
%
%
%
%
%
%
%
%
%
%

\begin{table}
\caption{Weak lensing errors only. $N_{\mathrm{FP}} = 1$}
\label{soloweaktable1}
\centering
\begin{tabular}{c c c} 
\hline\hline
 $n\,({\rm arcmin}^{-2})$ & $\Delta M/M$ & $\sigma(M)/M$ \\
 \hline
30 & 0.010 & 0.32 \\
50 & 0.002 & 0.26 \\
70 & -0.007 & 0.23 \\
\hline
\end{tabular}
\tablefoot{Relative mean error and standard deviation of the measured value of the total mass of the lens, for three different values of number density of background galaxies, $n$. }
\end{table}

\begin{table}
\caption{Weak lensing errors only. $N_{\mathrm{FP}} = 20$}
\label{soloweaktable2}
\centering
\begin{tabular}{c c c}
\hline\hline
 $n\,({\rm arcmin}^{-2})$ & $\Delta M/M$ & $\sigma(M)/M$ \\
 \hline
30 & 0.010 & 0.069 \\
50 & 0.008 & 0.060 \\
70 & 0.006 & 0.055 \\
\hline
\end{tabular}
\end{table}

\subsubsection{The realistic case}\label{realisticcase}

Finally we report the results of the simulations performed in the most general case, in which realistic conditions for both the weak lensing and the Fundamental Plane measurements are simulated.
In this case the simulations are performed for $n = 30,\,50,\,70\, {\rm arcmin}^{-2}$ and $N_{\mathrm{FP}} = 20$.
In Table \ref{tabellafinale} the results relative to 1000 runs are reported.
\begin{table}
\caption{General case. $N_{\mathrm{FP}} = 20$}
\label{tabellafinale}
\centering
\begin{tabular}{c c c}
\hline\hline
 $n\,({\rm arcmin}^{-2})$ & $\Delta M/M$ & $\sigma(M)/M$ \\ 
\hline
30 & 0.015 & 0.22  \\
50 & 0.003 & 0.22  \\
70 & 0.011 & 0.21  \\
\hline
\end{tabular}
\end{table}

The results confirm the picture emerged in \ref{soloweaklensing}: weak lensing errors play little role in determining the final dispersion on the estimate of the total mass.
In fact, the observed dispersion is practically equal to the one obtained in (\ref{stimaerrore}) by assuming perfect weak lensing measurements.
This test, although limited to a single lens model, confirms that it is possible, with a sufficient number of Fundamental Plane measurements uniformly distributed on the image plane, to break the mass--sheet degeneracy, at least for Coma cluster--like lenses at intermediate redshift.

\begin{figure}[!h]
\resizebox{\hsize}{!}{\includegraphics{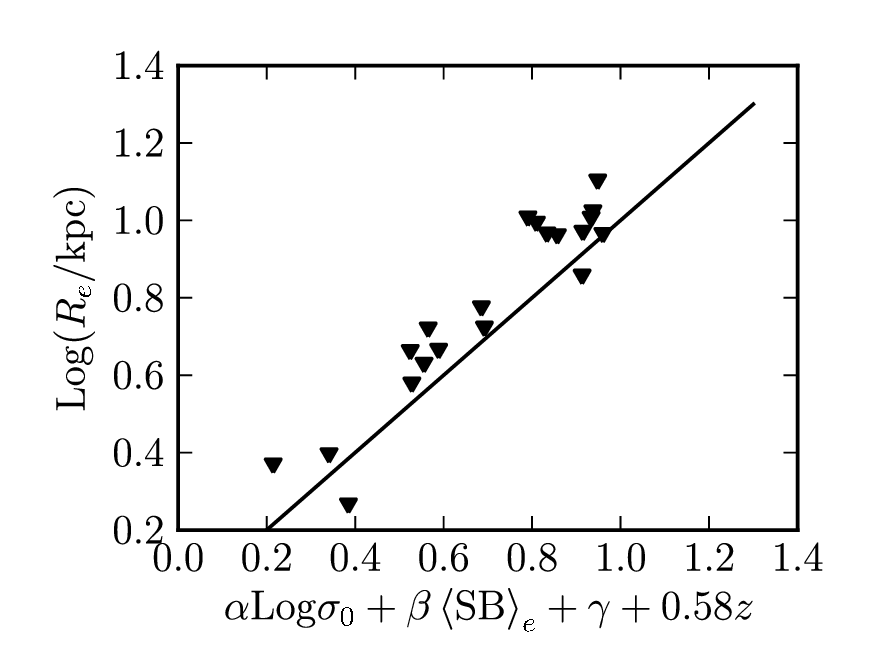}}
\caption{$N_{\mathrm{FP}} = 20$ simulated FP measurements, plotted in FP space. The straight line is the Fundamental Plane expected in the absence of lensing, viewed edge-on. Positions in the FP space are corrected for the evolution of the FP with redshift, following Treu et al. (\cite{Treu05}).}\label{FPshift}
\end{figure}

Finally, to better illustrate the effect of a cluster on lensed early--type galaxies, we show in Fig. \ref{FPshift} a set of simulated FP measurements, compared with the Fundamental Plane relation expected in the absence of lensing.
$N_{\mathrm{FP}} = 20$ objects in the redshift interval $0.5 < z < 1.0$ are generated and placed randomly behind the same cluster lens used for the previous simulations.
In constructing this plot, we assumed local values of the Fundamental Plane coefficients ($\alpha = 1.25$, $\beta = 0.32$, $\gamma = -8.970$ (J\o rgensen et al. \cite{Jorgensen96}, recomputed for $H_0 = 65 \mbox{ km s}^{-1}\mbox{ Mpc}^{-1}$ by Treu et al. \cite{Treu05}), corrected for evolution following Treu et al. (\cite{Treu05}): $\gamma(z) = \gamma(0) + 0.58z$.
The scatter of the FP is assumed to be 20\% in $r_e$.
The signature of the lensing signal can be clearly seen as an upward shift in the FP space.

\section{Conclusions and discussion}\label{Conclusions}

In this paper we presented a new lensing--based method for the measurement of the mass of galaxy clusters.
This method relies on the joint use of weak lensing data and magnification information, where the latter is obtained from Fundamental Plane measurements on background early-type galaxies.

A statistical study of the method was carried out, and simulations were performed to test the importance of the presence of substructures and of errors in the weak lensing analysis for the success of the measurement.

Our main conclusions are the following:
\begin{itemize}
\item The most important quantity on which the effectiveness of the method depends is the mean surface mass density within the field of view of observation, $\bar{\kappa}$, while little role is played by the shape of the mass distribution.
\item Substructures contribute at most with a scatter of a few percent on individual magnification measurements.
\item Weak lensing errors introduce only a small dispersion on the final estimate of the total mass.
\end{itemize}

On the basis of these results, we will now discuss which are the best lens candidates for an application of the present technique.

An important limit to the applicability of this method is the difficulty in performing Fundamental Plane measurements, since they require a significant amount of telescope time.
A realistic number of Fundamental Plane measurements that can be performed in an observational campaign is $\sim 20$.
Given this fact, we can fix $N_{\mathrm{FP}} = 20$ and discuss which systems are best analyzed with this number of magnification measurements.

One of the most important factors in determining whether a cluster can be realistically studied with our technique or not is its redshift.
The redshift must be be sufficiently high for the critical density to be low enough, to allow for higher values of $\kappa$ for a given physical surface mass density.
On the other hand, the redshift must also be sufficiently low so that it is possible to find an acceptable number of early-type galaxies behind it for which Fundamental Plane measurements can be performed.
The current observational capabilities and the lack of a calibration of the Fundamental Plane relation at very high redshifts set $z\sim 1.0$ as the highest redshift for which these measurements can be performed today.

\begin{figure}[!h]
\begin{center}
\resizebox{\hsize}{!}{\includegraphics{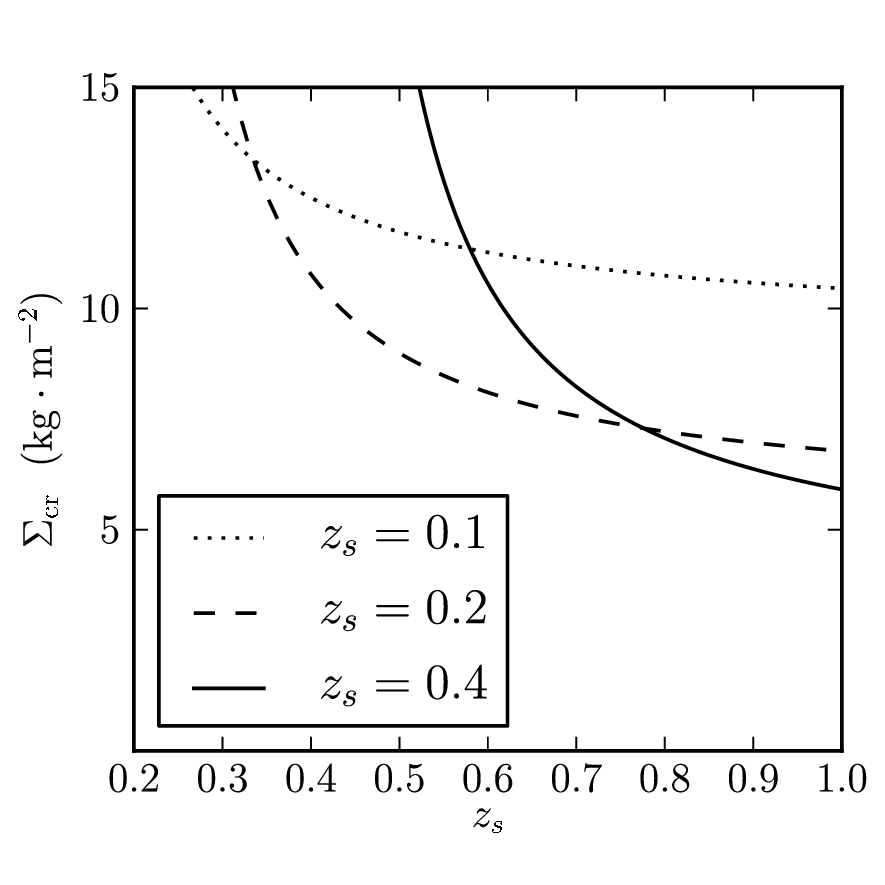}}
\caption{\small Critical density (in $\mbox{ Kg m}^{-2}$) as a function of source redshift for three different values of the lens redshift: $z_d = 0.1,\,0.2,\,0.4$.}\label{idealredshift}
\end{center}
\end{figure}

Bearing this in mind, we plot in Figure \ref{idealredshift} the value of the critical density as a function of source redshift for three different values of the lens redshift.
It can be seen that with a lens redshift $z_d = 0.1$ the resulting critical density is significantly higher than in the other cases at the source redshifts of interest and for this reason this case should be discarded.
At the opposite end, for a lens redshift $z_d =0.4$ the critical density is indeed the smallest for source redshifts higher than $\sim 0.8$, but the range of source redshifts for which the critical density is significantly small is limited to $z_s > 0.6$, and it shrinks rapidly for increasing $z_d$.
On the basis of these simple considerations we conclude that a suitable redshift range for our lens cluster is $z_d \sim 0.2\div 0.4$.
A thorough analysis of the problem would require a detailed knowledge of the redshift distribution of the observable early-type galaxies.

Then we can ask which intrinsic physical characteristics a cluster should have to be efficiently probed with Fundamental Plane measurements.

If we assume, conservatively, a mean dispersion of the Fundamental Plane relation of $20\%$ in $r_e$, it can be shown from (\ref{dispersionM2}) that in order to obtain a mass estimate with a precision of $20\%$ or better with $N_{\mathrm{FP}} = 20$ Fundamental Plane measurements with uniform distribution in redshift between $z=0.5$ and $z=1.0$, the average surface mass density of the lens should be $\bar{\kappa} \gtrsim 0.3$.

Clearly, the quantity $\bar{\kappa}$ depends on the mass distribution of the lens but also on the size of the field of observation, which is set by the observer. How critical is the choice of this latter quantity for determining a value $\bar{\kappa} =0.3$?
One can always restrict the observations to the inner regions of a given cluster, to increase $\bar{\kappa}$.
However, there is a limit set by the number density of background early--type galaxies for which Fundamental Plane measurements can be effectively performed.
This number density was estimated in \cite{BL06} to be $\sim 2\,\mathrm{arcmin}^{-2}$. 
%
%
%
%
%
%
%
%
%
%
%
%
%
%
%
%
%
%
Therefore, if we wish to find $N_{\mathrm{FP}} = 20$ objects suitable for our purposes, we need to cover a field of view of 10 ${\rm arcmin}^2$.

At this point we can study what is the minimum mass a cluster should have in order to have a mean surface mass density $\bar{\kappa} = 0.3$ within a circle of area $A = 10\,{\rm arcmin}^2$ (and radius $\theta^* = \sqrt{10/\pi}\simeq 1.8\,{\rm arcmin}$).
Fixing the lens redshift $z_d = 0.3$, this value of the surface mass density within the circle corresponds to a value of the enclosed projected mass equal to $M_{\mbox{enc}} = 4.2\times10^{14}\,M_\odot$.
For a NFW profile with concentration parameter $c=10$, this value corresponds to a limiting virial mass
\begin{equation}\label{massalimite}
M_{200}^* = 8.0\times10^{14}\,M_\odot,
\end{equation}
and similar values hold for different values of the concentration $c$.
The quantity $M_{200}^*$ is the minimum mass a NFW cluster should have in order to satisfy $\bar{\kappa} \geq 0.3$ within a circle of area $A = 10\,{\rm arcmin}^2$ in the sky.
In that case the mass of the cluster within the circle can be measured with a 20\% precision or better with 20 Fundamental Plane measurements and a weak lensing analysis.

If we want to extend the analysis to the outer parts of our cluster, we must increase the number of Fundamental Plane measurements, to compensate for the reduced value of the average surface mass density $\bar{\kappa}$.

If we want to improve the sensitivity to less massive clusters, we must either increase the number of Fundamental Plane measurements for fixed aperture or choose a smaller aperture for fixed $N_{\mathrm{FP}}$.
In any of these cases, a higher number density of sources is needed, which cannot be obtained without an improvement of the observational capabilities.
This shows that the number density of observable sources is indeed a key factor and confirms that the lens redshift should not be too high.

A possible alternative strategy is to select only the brighter objects, since it has been recognized that the FP scatter decreases with increasing mass (Treu et al. \cite{Treu05}; van der Wel et al \cite{vanderWel}): this would allow for a more precise estimate of the magnification for a given lens  and fixed number of FP measurements.
The feasibility of such an approach also depends critically on the number density of background sources.

On the basis of this discussion, we conclude that the value of $M_{200}^*$ given by (\ref{massalimite}) is an estimate of the minimum mass a cluster should have to allow for a mass measurement with the present method.
The value is on the high side, but there are indeed many clusters that have observed values of $M_{200}$ higher than this threshold.
Well-studied examples are A1689, A1703, A370, RX J1347-11 (see Broadhurst et al. \cite{Broad2008} for a review).
All these systems display strong lensing features that allow for a good estimate of the mass distribution in the inner ($\theta < 1\,{\rm arcmin}$) regions of the clusters.
One might think that the availability of strong lensing data would rule out the need for other observations in such clusters.
Nevertheless, a great benefit would come from the addition of Fundamental Plane measurements as they could provide important constraints on the mass distribution within a radius at least 2 times larger.

In summary, the method presented in this paper is, because of its nonparametric form, a potentially powerful tool to break the mass--sheet degeneracy in lensing studies of clusters of galaxies.
As shown in Sect.~\ref{statistics}, it allows also for a relatively easy estimate of the accuracy of the total mass measurement. 

In addition, the method can be extended by allowing for the inclusion of magnification measurements obtained from different means, for example from the observation of type Ia supernovae (see, e.g., Holz \cite{Holz}; Goobar et al. \cite{Goobar}; J\"{o}nsson et al. \cite{Jonsson}).
The statistical framework developed in this paper can be applied with little effort to such more general situations, and therefore can be used as a reference framework to estimate the degree of precision of other methods that rely on the combination of weak lensing and magnification measurements.

\begin{appendix} 
\section{Probability distribution of the estimates of $\kappa_{\mathrm{FP}}$}\label{appendixA}
Here we will discuss the choice of a Gaussian probability distribution for the estimates $\kappa_{\mathrm{FP}}$ of the surface mass density obtained by combining Fundamental Plane and weak lensing measurements.
The observable quantity is the ratio $r \equiv r_e^{\mathrm{(FP)}}/r_e^{\mathrm{(obs)}}$ of the galaxy effective radius inferred from the Fundamental Plane relation to the observed (magnified) effective radius.
Of these two quantities, the first, with its 15\% dispersion, is by far the one with the larger uncertainty.
Thus, if the probability distribution of the estimate of $r_e^{\mathrm{(FP)}}$ is Gaussian, the ratio $r$ will have a nearly Gaussian distribution as well.
Let us assume this is the case.

The surface mass density $\kappa^{\mathrm{(FP)}}$ at the image position can be expressed as a function of magnification $\mu$, galaxy redshift $z$ and average distortion $\left<\epsilon\right>$, as in (\ref{kappaFP}), (\ref{abkappa}), (\ref{ckappa}).
Let us assume that no error comes from the measurements of $z$ and $\left<\epsilon\right>$, in accordance with the hypotheses of Sect.~\ref{statistics}.
The only error source is then the measurement of the magnification.
Note that, if $\kappa^{\mathrm{(FP)}}$ depends linearly on the ratio $r$, then $\kappa^{\mathrm{(FP)}}$ would also have a Gaussian probability distribution.
However, the dependence of $\kappa^{\mathrm{(FP)}}$ on magnification is more complicated than a simple linear relation.
In fact, $\kappa^{\mathrm{(FP)}}$ is given by
\begin{equation}\label{kappadaratio}
\kappa^{\mathrm{(FP)}} = \frac{-b-\sqrt{b^2 - a\left(c - r^2\right)}}{a}.
\end{equation}
Now, if the condition
\begin{equation}\label{condition}
b^2 - ac \ll ar^2
\end{equation}
holds, then the square root can be approximated by
\begin{equation}
\sqrt{b^2 - ac +ar^2} \approx \sqrt{a}r
\end{equation}
and (\ref{kappadaratio}) becomes linear.
As a consequence, the Gaussianity of $r$ translates into the Gaussianity of $\kappa^{\mathrm{(FP)}}$.
At this point, we need to demonstrate the validity of (\ref{condition}).
The ratio $r$ is typically of order 1, unless in the case of large magnifications.
Thus, it is sufficient to prove that $b^2 - ac \ll a$. 
We begin by noting that $a \sim Z_i^2[1 - O(|\left<\epsilon\right>|^2)]$. The quantity $|\left<\epsilon\right>|$ is typically small, except in the proximity of critical curves, while the combination of weights that multiplies $|\left<\epsilon\right>|^2$ in $a$ is of order unity.
Similarly, $b \sim Z_i[1 - O(|\left<\epsilon\right>|^2)]$ and $c \sim 1 - O(|\left<\epsilon\right>|^2)$. Then, the quantity $b^2 - ac$ is of order $Z_i^2O(|\left<\epsilon\right>|^2) \ll 1$ .
Since the right hand side of (\ref{condition}) is of order unity, (\ref{condition}) is satisfied.
This implies that the dependence of $\kappa^{\mathrm{(FP)}}$ on $r$ is approximately linear.
Then, if the probability distribution of $r$ is Gaussian, which is reasonable, the same applies to $\kappa^{\mathrm{(FP)}}$.

\end{appendix}

\end{document}